\RequirePackage[T1]{fontenc}
\documentclass[aps,pra,reprint,superscriptaddress,nofootinbib,floatfix]{revtex4-1}
\usepackage{graphicx}
\usepackage{amssymb}
\usepackage{mathtools}
\usepackage{braket}
\usepackage{bbold}

\DeclareMathOperator{\cov}{cov}

\usepackage{xcolor}
\usepackage{float}
\usepackage[shortcuts]{extdash}

\usepackage{siunitx}

\definecolor{bucketcolour}{RGB}{255,153,0} 
\definecolor{NRcolour}{RGB}{153, 0, 204} 
\definecolor{bucketTRcolour}{RGB}{255,153,0} 
\definecolor{NRTRcolour}{RGB}{153, 0, 204} 
\definecolor{noHOMcolour}{RGB}{143, 252, 255} 

\usepackage{acro}
\acsetup{single}
\DeclareAcronym{CFI}{
	short = CFI,
	long = classical Fisher information,
}
\DeclareAcronym{CRB}{
	short = CRB,
	long = Cram{\'e}r-Rao bound,
}
\DeclareAcronym{FI}{
	short = FI,
	long = Fisher information
}
\DeclareAcronym{FIM}{
	short = FIM,
	long = Fisher information matrix
}
\DeclareAcronym{HOM}{
	short = HOM,
	long = Hong-Ou-Mandel,
}
\DeclareAcronym{MLE}{
	short = MLE,
	long = maximum likelihood estimator
}
\DeclareAcronym{QCRB}{
	short = QCRB,
	long = quantum Cram{\'e}r-Rao bound,
}
\DeclareAcronym{QFI}{
	short = QFI,
	long = quantum Fisher information,
}
\DeclareAcronym{SPDC}{
	short = SPDC,
	long = spontaneous parametric down-conversion,
}
\DeclareAcronym{POVM}{
	short = POVM,
	long = positive-operator valued measure,
}
\DeclareAcronym{SNSPD}{
	short = SNSPD,
	long = superconducting nanowire single-photon detector,
}
\DeclareAcronym{TES}{
	short = TES,
	long = transition-edge sensor,
}

\usepackage{hyperref}
\hypersetup{
    colorlinks=true,
    citecolor=blue,
    filecolor=blue,
    linkcolor=blue,
    urlcolor=blue
}

\newcommand{\estimator}[1]{\check{#1}}

\renewcommand{\vec}[1]{\boldsymbol{#1}}

\newcommand{\Htwoph}{H_{\delta}^{\mathrm{2ph}}}
\newcommand{\Irel}{I_{\mathrm{rel}}}
\newcommand{\omegap}{\omega_{\mathrm{p}}}

\NewDocumentCommand{\DB}{s+m}{\textcolor{purple}{\IfBooleanT{#1}{\bfseries} #2}}
\NewDocumentCommand{\HS}{s+m}{\textcolor{cyan}{\IfBooleanT{#1}{\bfseries} #2}}
\NewDocumentCommand{\EG}{s+m}{\textcolor{blue}{\IfBooleanT{#1}{\bfseries} #2}}
\NewDocumentCommand{\NW}{s+m}{\textcolor{magenta}{\IfBooleanT{#1}{\bfseries} #2}}

\begin{document}
\title{Beyond coincidence: using all the data in Hong-Ou-Mandel interferometry}

\author{Hannah Scott}
\affiliation{SUPA, Institute of Photonics and Quantum Sciences, Heriot-Watt University, Edinburgh, EH14 4AS, UK}
\author{Dominic Branford}
\affiliation{SUPA, Institute of Photonics and Quantum Sciences, Heriot-Watt University, Edinburgh, EH14 4AS, UK}
\author{Niclas Westerberg}
\affiliation{School of Physics and Astronomy, University of Glasgow, Glasgow, G12 8QQ, United Kingdom}
\affiliation{SUPA, Institute of Photonics and Quantum Sciences, Heriot-Watt University, Edinburgh, EH14 4AS, UK}
\author{Jonathan Leach}
\affiliation{SUPA, Institute of Photonics and Quantum Sciences, Heriot-Watt University, Edinburgh, EH14 4AS, UK}
\author{Erik M. Gauger}
\affiliation{SUPA, Institute of Photonics and Quantum Sciences, Heriot-Watt University, Edinburgh, EH14 4AS, UK}

\date{\today}

\begin{abstract}
The Hong-Ou-Mandel effect provides a mechanism to determine the distinguishability of a photon pair by measuring the bunching rates of two photons interfering at a beam splitter.
Of particular interest is the distinguishability in time, which can be used to probe a time delay.
Photon detectors themselves give some timing information, however---while that resolution may dwarf that of an interferometric technique---typical analyses reduce the interference to a binary event, neglecting temporal information in the detector.
By modelling detectors with number and temporal resolution we demonstrate a greater precision than coincidence rates or temporal data alone afford.
Moreover, the additional information can allow simultaneous estimation of a time delay alongside calibration parameters, opening up the possibility of calibration-free protocols and approaching the precision of the quantum Cram{\'e}r-Rao bound.
\end{abstract}

\maketitle

\section{Introduction}\label{sec_intro}

When a pair of indistinguishable photons enter the two input ports of a balanced beamsplitter, they will always be detected in the same (randomly selected) output mode.
This ``bunching'' of photons entails a reduction of coincident clicks between two detectors which monitor the output ports.
A complete lack of detected coincidences indicates that an incoming pair of photons
is perfectly indistinguishable in all degrees of freedom, including spatial, frequency, and polarisation properties; and most importantly to this work, the photons must also possess an exactly matched temporal profile.
More generally, the coincidence rate constitutes a reliable and straightforward way of probing the difference between two pure photonic states.
This is the effect the eponymous Hong, Ou, and Mandel demonstrated in their seminal work from 1987~\cite{Hong-1987}.

Since then, the \ac{HOM} effect has become a key tool in quantum optics~\cite{Boto-2000,Kok-2007,Ma-2012,Ferraro-2015,Giovannini-2015,Kim-2016,Agnesi-2019}.
Most relevantly to this paper, it is used in a variety of situations throughout quantum metrology: for parameters that cannot be easily or efficiently measured directly, it provides a way of estimating them through their impact on the indistinguishability of a photon pair.
This is traditionally used as a way of measuring very small time delays~\cite{Olindo-2006,Lyons-2018,Chen-2019,Yang-2019,Restuccia-2019}, but can include other parameters such as polarisation~\cite{Harnchaiwat-2020}. A related, albeit slightly different quantum optical method for measuring path delays is quantum optical coherence tomography~\cite{Nasr-2003,Mazurek-2013,Taylor-2015}.

High precision time delay measurements have a variety of scientific and technological applications.
It has been proposed that the imaging of delicate (biological) samples can benefit from exploiting quantum effects to limit photodamage~\cite{Wolfgramm-2013,Taylor-2016, Casacio-2020, Garces-2020}.
For instance, to resolve the thickness of an unknown sample, such as e.g.~a cell membrane, a delay can be induced in the path of a photon by allowing it to pass through the sample, accompanied by a second photon which does not pass through the sample.
The simplest, and perhaps most obvious, way to measure this delay would be to place detectors at the end of both photons' paths and measure the difference in their respective arrival times, an approach similar to a direct time-of-flight measurement~\cite{Pellegrini-2000,tobin-2017}.
However, the time-resolution of conventional single-photon detectors is orders of magnitude too poor to capture the precision needed for these measurements to provide the desired nanoscale resolution.

By contrast, \ac{HOM} metrology has recently achieved attosecond resolution~\cite{Lyons-2018}.
In this case, rather than measuring the delay directly, changes to the coincidence counts at two detectors placed at the outport ports of the HOM beamsplitter are recorded and analysed.
Assuming an initial calibration with regards to all the other factors influencing this indistinguishability, the time delay can be isolated and estimated.
Traditionally, this achieves sub-femtosecond resolution by scanning through and observing a shift of the entire HOM dip~\cite{Giovannini-2015}, but more more sophisticated protocols can perform up to $100\times$ better~\cite{Lyons-2018}.

This attosecond resolution goes well beyond that of the single-photon detectors which may be used in such an experiment%
\footnote{For instance, the conventional avalanche photodiode  of Ref.~\onlinecite{SPCM-AQRH} has has a temporal resolution of \SI{\approx 350}{\ps}, 
but some recent developments in detector time-resolution are approaching picosecond precision~\cite{Rath-2016,Zadeh-2018,Zhu-2019,Korzh-2020}.} limited to the nanosecond--picosecond range (centimetre-scale).
Even allowing for the additional complexity of a HOM interferometer, this chasm seems hard to overcome with a direct timing approach
Nonetheless, we need not overlook entirely the time-resolving capabilities of single-photon detectors.
Currently their time-resolution is only used for the purpose of establishing coincidence matching time windows.
As we shall see, even when their precision falls short of that of HOM interferometry, making better use of their temporal information is advantageous in multiple ways.
\begin{figure*}[htb]
\centering
\includegraphics[width=\linewidth]{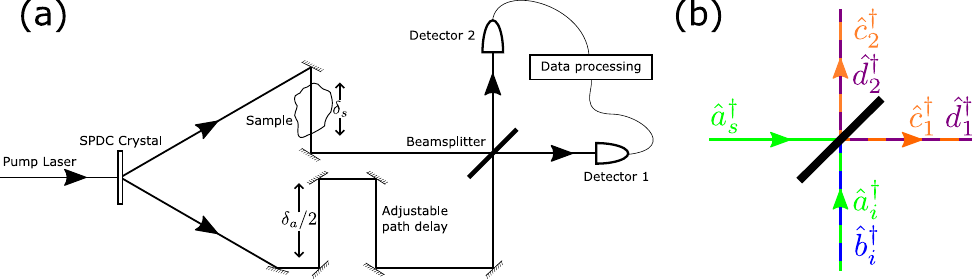}
\caption{(a) A schematic of our protocol.
Photons are generated via \acl{SPDC}.
In the top arm, the photon passes through a sample and is delayed by $\delta_s$.
An adjustable delay $\delta_a$ is placed in the lower arm.
Both photons then interfere at the beamsplitter, with a relative time delay $\delta = \delta_s-\delta_a$.
Detectors D1 and D2 are placed behind the beamsplitter output ports.
Each detection event is then assigned to a time bin and checked for coincidence or bunching.
(b) How we model the photon modes either side of the beamsplitter in Sec.~\ref{sec_prob_analysis}.
In the lower arm, an additional orthogonal mode $\hat{b}^\dagger$ is present to capture differences other than temporal between the photons 
A mixture of modes $\hat{c}^\dagger$ and $\hat{d}^\dagger$ leave both beamsplitter ports.}
\label{PD_fig1}
\end{figure*}

In this Article we analyse the benefits of incorporating detector time resolution capabilities into HOM metrology.
We find that not only does this upgrade the performance of HOM interferometers, but it could also facilitate a simplified protocol by eliminating the need for a separate calibration stage.
Furthermore, recent developments in the feasibility of number-resolving detectors~\cite{Kardynal-2008,Young-2019,Zhu-2019}, motivate us to also analyse the benefits of photon number-resolution, and we find that these would also lead to a tangible performance enhancement.
We show that by combining number and time resolving detectors the fundamental \ac{QCRB} can be approached with a \ac{HOM} sensing protocol.

This Article is organised as follows: We start in Sec.~\ref{sec_protocol_description} by explaining our \ac{HOM} protocol and the model of time-resolution. Section~\ref{sec_prob_analysis} details the fundamental probability analysis behind our protocol, and relates it to our parameter estimation procedure in Sec.~\ref{sec_para_est} through the Cram{\'e}r-Rao bound. In Sec.~\ref{sec_ideal_detectors}, we derive the \ac{QFI} for the time delay, giving the ceiling on performance for any measurement protocol, and outline one such (technically infeasible) protocol. Our main results are presented in Sec.~\ref{sec_results}, to show where our proposed protocol would sit between this ceiling and previous protocols without time- or number-resolving capabilities. Finally, we discuss our findings in Sec.~\ref{sec_discussion}, outlining some of the limitations of our analysis, explaining and justifying the assumptions and simplifications made, and summarising our key conclusions.

\section{Protocol description}\label{sec_protocol_description}

We consider a standard HOM protocol~\cite{Branczyk-2017} as schematically depicted in Fig.~\ref{PD_fig1},  but our analysis will include detectors with time- and number-resolving capabilities.
The protocol begins with a frequency-entangled pair of photons propagating along two spatial input modes of equal path length to the \ac{HOM} beamsplitter.
Typically, the photon pair is generated by a pump laser in a nonlinear crystal via \acf{SPDC}.
One of the photons traverses the sample and picks up some small path length and associated time delay $\delta_s$. 
The other path contains a precisely adjustable delay $\delta_a$.
Estimating $\delta_s$ could now be accomplished by scanning the adjustable delay to resolve the entire \ac{HOM} dip with and without sample present and evaluating its shifts. 
Alternatively, we could tune $\delta_a$ such as to give minimal coincidence counts in both cases (i.e.~place ourselves at the bottom of the dip) and read off the difference. 
Neither approach is optimal and we here follow the idea of Refs.~\cite{Lyons-2018, Harnchaiwat-2020} to coarse-tune $\delta_a$ such that the relative delay ${\delta = \delta_s - \delta_a}$ maximises the information we obtain for each input photon pair. 
In most practical situations this point is near the inflection point of the dip and $\delta_a \gg \delta_s$. 
Now, comparing the difference in coincidence rate with and without sample present allows the estimation of $\delta_s$. In the protocol employed by Refs.~\cite{Lyons-2018, Harnchaiwat-2020}, the maximum likelihood estimator for $\delta_s$ requires a fit to the previously recorded \ac{HOM} dip of the setup.  
In those cases the employed fit was based on an inverted Gaussian and required determination of the visibility (depth of the dip) and the dip width as two calibration parameters, as well as knowledge about the photon loss rate that emerges from detector efficiency and other experimental imperfections.

\begin{figure*}[htb]
\centering
\includegraphics[width=\linewidth]{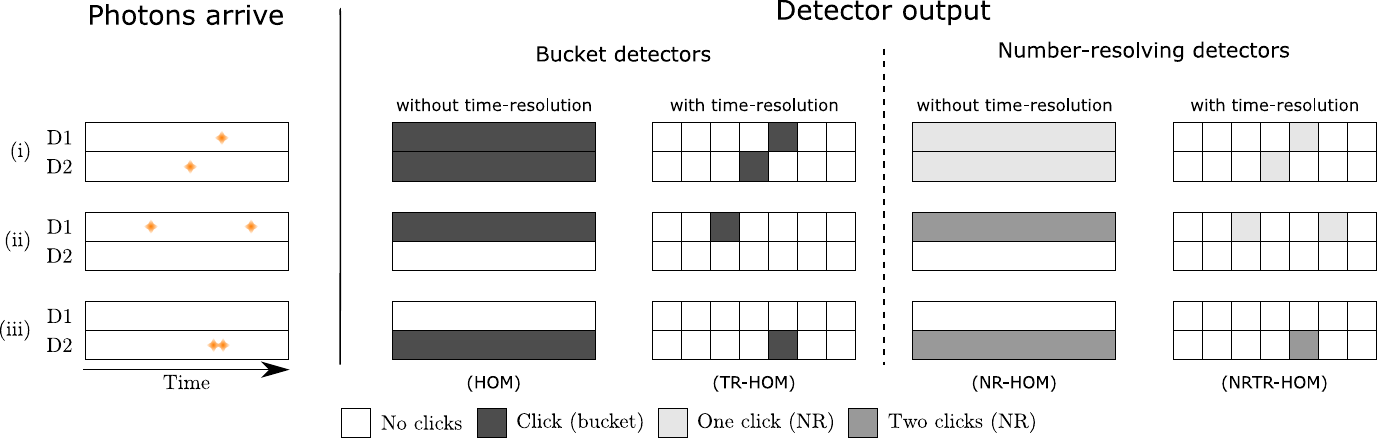}
\caption{Possible outputs for a pair of photons incident on the different types of detector we consider. First, standard (non-time-resolving) bucket detectors: each clicks only once whether one or more photons arrive. Then, bucket detectors with time-resolving capabilities: each detector click is now associated with a specific time bin, giving some precision as to when a photon arrives. As a consequence of detector dead time if two photons arrive at the same detector then it only clicks once, for the time bin when the earlier photon arrived. Then, number-resolving detectors without time-resolving capabilities: the detector can now differentiate between one or two photons arriving. Finally, number-resolving detectors with time-resolving capabilities: now, it can indicate when each photon arrived, up to the precision of our time bins.}
\label{PD_fig2}
\end{figure*}

We shall analyse the potential performance of \ac{HOM} time delay estimation protocols for four different kinds of detectors whose behaviour is illustrated in Fig.~\ref{PD_fig2}.
We begin with distinguishing \textit{bucket detectors}, that are unable to discriminate the number of photons triggering a click
(assuming the detector dead-time is sufficiently long such that a single detector can deliver at most one click per bunched photon pair),
from \textit{number-resolving detectors}, which can reliably provide two clicks whenever they are hit by two bunched photons. 
In both cases we further consider detectors capable of providing a reliable and precise timestamp to the clicks as well as those whose signal is so smeared out that it contains little useful timing information. 
Previous \ac{HOM} protocols have typically utilised the latter kind (or ignored available timing information for the delay estimator)\cite{Lyons-2018, Giovannini-2015, Lyons-2018_2}. 

Experimental scenarios do not involve just a single photon pair and the pump laser is either in CW or pulsed operation.
In the former case, the \ac{SPDC} events happen randomly in time with the average duration between them determined by the pump intensity, in the latter we have a probability of generating one, or more, pairs with each pulse separated in time by the repetition frequency. 
In either case, \ac{HOM} protocols rely on defining a coincidence time window (which is smaller than the expected time between photon pairs) to perform coincidence matching. 
However, as pairs can be spaced out as needed, this does not place stringent demands on the detection time-resolution capacity and is compatible with significant amounts of timing jitter (typical coincidence windows last on the order of nanoseconds). 
If this is the only sense in which detector timestamps are utilised, we shall refer to this scenario as detection \textit{without time-resolution}.

By contrast, we model detection \textit{with time-resolution} by sorting detected photons into equally-sized detection time bins of temporal width $T$, which reflect the timing capability and fidelity of the detection hardware.
A visualisation of this time binning process, and the resultant output available, is shown in Fig.~\ref{PD_fig2} for the four different types of detector we consider. Based on these outputs, we are able to estimate the relative delay $|\delta|$\footnote{As we cannot track which photon is which after the beamsplitter, sign information is lost and we can only truly estimate $|\delta|$, the size of the delay. In practice this is unimportant, but is worth remarking on as this technicality does not occur in protocols based purely on arrival time measurements, as there no beamsplitter is present.}.


\section{Probability analysis}\label{sec_prob_analysis}

Having established the nature of the data available for the time delay estimation, we now proceed with establishing the probabilities for recording this data as a function of the time delay, loss processes and detector efficiency, and other sources distinguishability between our photon pair. 
We begin this calculation with the typical initial \ac{SPDC} process.

\subsection{Biphoton state}
A biphoton state with fixed total energy, generated by \ac{SPDC}, takes the standard form~\cite{Hong-1987,Grice-1997}
\begin{equation}
	\ket{\psi} = \int_0^{\omegap} d\omega \> \phi(\omega) \hat{a}_i^\dagger(\omega) \hat{a}_s^\dagger(\omegap-\omega) \ket{0}
	\label{PA_eqn1}
\end{equation}
with $\omegap$ the pump frequency; and the operators $\hat{a}_i^{\dagger}(\omega)$ and $\hat{a}_s^{\dagger}(\omegap-\omega)$ creating the two photons with frequencies $\omega$ and $\omegap - \omega$, respectively.
In the above, $\phi(\omega)$ is the joint spectral amplitude, which we take to be Gaussian and of the form
\begin{equation}
	\phi(\omega) = (2\pi\sigma^2)^{-1/4} \> e^{-\frac{(\omega-\omegap/2)^2}{4\sigma^2}} ~,
	\label{PA_eqn2}
\end{equation}
where $\sigma$ is the spectral width~\cite{Branczyk-2017} and $1/\sigma$ corresponds to the coherence time of the photons.  The spectral width of the biphoton state therefore directly impacts the temporal width of the dip.
By assuming that our distribution of frequencies is strongly peaked around $\omegap/2$ we can extend our $\omega$ integration over the range $(-\infty,\infty)$ to simplify our calculations.

In practice, however, the photons will not be generated in modes which are---up to the spatial mode---entirely indistinguishable.
We model this indistinguishability by the introduction of an orthogonal mode $\hat{b}_i$ in the lower arm%
\footnote{The orthogonal component can be localised to one mode without loss of generality by labelling the top mode \( \hat{a}_s \), regardless of whether it is identical to the desired or intended mode.}.
We thus rewrite
\begin{equation}
	\hat{a}_i^\dagger(\omega)\to\sqrt{\alpha}\>\hat{a}_i^\dagger(\omega)+\sqrt{1-\alpha}\>\hat{b}_i^\dagger(\omega)
	\label{PA_eqn17}
\end{equation}
with the parameter $\alpha$ describing the relative indistinguishability of our photons; $\alpha \in [0,1]$ encodes the {\it visibility}, i.e.~the `depth' of the HOM dip, for simplicity we take the relative phase between \( \hat{a}_i \) and \( \hat{b}_i \) to be real, this phase does not affect the resulting probabilities.
The state just before the beamsplitter now takes the form
\begin{equation}
\begin{aligned}
	\ket{\psi} &= \int d\omega \> \phi(\omega) e^{-i\omega\delta_a} e^{-i(\omegap-\omega)\delta_s}  \\*
	& \times [\sqrt{\alpha}\>\hat{a}_i^\dagger(\omega)+\sqrt{1-\alpha}\>\hat{b}_i^\dagger(\omega)]\hat{a}_s^\dagger(\omegap-\omega)\ket{0}~.
\end{aligned}
	\label{PA_eqn18}
\end{equation}

As the photons travel towards the beamsplitter, we are interested in the relative time delay ${\delta=\delta_s-\delta_a}$. In which case we work with the state
\begin{equation}
\begin{aligned}
	\ket{\psi} &= \int d\omega \> \phi(\omega) e^{i\omega\delta} \\*
	&\mkern16mu \times [\sqrt{\alpha}\>\hat{a}_i^\dagger(\omega)+\sqrt{1-\alpha}\>\hat{b}_i^\dagger(\omega)]\hat{a}_s^\dagger(\omegap-\omega)\ket{0}~,
\end{aligned}
	\label{PA_eqn99}
\end{equation}
where we ignore the irrelevant global phase of $e^{-i\omegap \delta_s}$.
The \ac{HOM} beamsplitter then transforms our operators as
\begin{gather}
	\hat{a}_i^\dagger(\omega) \to \frac{1}{\sqrt{2}}[i\hat{c}_1^\dagger(\omega)+\hat{c}_2^\dagger(\omega)]~,\nonumber \\*
	\hat{b}_i^\dagger(\omega) \to \frac{1}{\sqrt{2}}[i\hat{d}_1^\dagger(\omega)+\hat{d}_2^\dagger(\omega)]~,\nonumber \\*
	\hat{a}_s^\dagger(\omega) \to \frac{1}{\sqrt{2}}[\hat{c}_1^\dagger(\omega)+i\hat{c}_2^\dagger(\omega)]~,
	\label{PA_eqn19}
\end{gather}
where the indices $1$ and $2$ denote the two output ports of our beamsplitter (or, correspondingly, which detector the photon then arrives at).
Our state finally becomes
\begin{align}
	\ket{\psi}&=\frac{1}{2}\int d\omega \> \phi(\omega) e^{i\omega\delta} (\sqrt{\alpha}[i\hat{c}_1^\dagger(\omega)+\hat{c}_2^\dagger(\omega)] \nonumber \\*
	& \qquad \qquad \qquad +\sqrt{1-\alpha}\>[i\hat{d}_1^\dagger(\omega)+\hat{d}_2^\dagger(\omega)]) \nonumber \\* 
	& \qquad \qquad \times [\hat{c}_1^\dagger(\omegap-\omega)+i\hat{c}_2^\dagger(\omegap-\omega)]\ket{0}~.
\label{PA_eqn20}
\end{align}

\subsection{Fundamental probabilities}
We proceed to determine the probabilities for photon bunching and anti-bunching (the latter corresponding to a coincidence at the detectors).
These are, respectively, cases where both photons arrive in the same detector with a delay of magnitude \( \tau \), and where both photons arrive in different detectors with a delay of magnitude \( \tau \).

These detection events%
\footnote{In calculating detection probabilities, note again that we are assuming $\phi(\omega)$ is strongly peaked around a central frequency of $\omega_p/2$.
This allows a standard simplification of the electric field and we therefore work in terms of the photon creation and annihilation operators~\cite[Chap.~6]{Loudon-2000}.
In this way, we treat a detector click as analogous to the arrival of a photon.}
are the probabilities of detecting 
\( 
\hat{c}_1^{\dagger} (t\pm\frac{\tau}{2}) \hat{c}_2^{\dagger} (t\mp\frac{\tau}{2}) \ket{0} , \)
\( \hat{c}_1^{\dagger} (t\pm\frac{\tau}{2}) \hat{d}_2^{\dagger} (t\mp\frac{\tau}{2}) \ket{0} , \)
and 
{\( \hat{d}_1^{\dagger} (t\pm\frac{\tau}{2}) \hat{c}_2^{\dagger} (t\mp\frac{\tau}{2}) \ket{0} , \)}
for coincidence events; and 
{\( \hat{c}_j^{\dagger} (t-\frac{\tau}{2}) \hat{c}_j^{\dagger} (t+\frac{\tau}{2}) \ket{0} \)}
and \( \hat{c}_j^{\dagger} (t\pm\frac{\tau}{2}) \hat{d}_j^{\dagger} (t\mp\frac{\tau}{2}) \ket{0} \)
(with \( j = 1,2 \)), for bunching events\footnote{Considering only those events involving photon pairs.} with \( t \) being the average arrival time.
From these we can construct a \ac{POVM} with elements
\( \hat{\Pi}_{\mathrm{c}} (\tau) \) and \( \hat{\Pi}_{\mathrm{b}} (\tau) \) 
corresponding to a coincidence or bunching detection with arrival times differing by \( \tau \) (see Appendix~\ref{POVMs} for more details).

We can now evaluate the coincidence and bunching probabilities for a given difference in arrival times $\tau$: ${P_\mathrm{c}(\tau)=\bra{\psi}\hat{\Pi}_\mathrm{c}(\tau)\ket{\psi}}$ and ${P_\mathrm{b}(\tau)=\bra{\psi}\hat{\Pi}_\mathrm{b}(\tau)\ket{\psi}}$, respectively.
Note that in this sense coincidence and bunching have lost the meaning of simultaneity and only refer to the spatial mode in which the photons are found (within the relatively long coincidence window).
We obtain
\begin{align}
	P_\mathrm{c}(\tau) &= \frac{\sigma}{\sqrt{2\pi}} e^{-2\sigma^2(\delta+\tau)^2} \left(1+e^{8\delta\sigma^2\tau}-2\alpha e^{4\delta\sigma^2\tau}\right)~,
	\label{PA_eqn24} \\
	P_\mathrm{b}(\tau) &= \frac{\sigma}{\sqrt{2\pi}} e^{-2\sigma^2(\delta+\tau)^2} \left(1+e^{8\delta\sigma^2\tau}+2\alpha e^{4\delta\sigma^2\tau}\right)~.
	\label{PA_eqn36}
\end{align}
Closer inspection shows that only at perfect visibility, $\alpha=1$, do the photons become wholly indistinguishable at $\delta=0$. 
The full derivation for these two expressions is detailed in Appendix~\ref{POVMs}.

By integrating over all $\tau$ we obtain the total coincidence and bunching rates, independent of the photons' arrival times, as usually seen in \ac{HOM} analyses~\cite{Branczyk-2017}:
\begin{align}
	P_{\mathrm{c},\mathrm{tot}} &= \int_{0}^\infty d\tau \> P_\mathrm{c}(\tau) = \frac{1}{2}\left(1-\alpha e^{-2\sigma^2\delta^2}\right)~,
	\label{PA_eqn25} \\
	P_{\mathrm{b},\mathrm{tot}} &= \int_{0}^\infty d\tau \> P_\mathrm{b}(\tau) = \frac{1}{2}\left(1+\alpha e^{-2\sigma^2\delta^2}\right)~.
	\label{PA_eqn26}
\end{align}

We now introduce the binning that captures the finite time resolution of detectors into our analysis. 
Returning to Eqs.~\eqref{PA_eqn24} and~\eqref{PA_eqn36}, which are valid for perfect timing, we collect all events that occur within the temporal width $T$ into the same indexed bin.
The difference between recorded bin indices then gives coarse-grained information about the delay $\tau$ between two subsequent detection events.

Our time bins are fixed in duration and our photons are equally likely to arrive at any point within the bin. 
For time bin width $T$, and $nT<\tau<(n+1)T$, the total probability of the two photons arriving $n$ bins apart is $[(n+1)T-\tau]/T$, and  probability of them arriving $n+1$ bins apart is $(\tau-nT)/T$.
We can then write the probability of a coincidence separated by $n$ time bins, $P_{\mathrm{c},n}$, as
\begin{align}
	P_{\mathrm{c},0} &= \left[\int_0^T d\tau \> \frac{T-\tau}{T}P_\mathrm{c}(\tau)\right]\nonumber \\*
	P_{\mathrm{c},n>0} &= \left[\int_{(n-1)T}^{nT} d\tau \> \frac{\tau-(n-1)T}{T}P_\mathrm{c}(\tau)\right.\nonumber \\*
	& \qquad \left. + \int_{nT}^{(n+1)T} d\tau \> \frac{(n+1)T-\tau}{T}P_\mathrm{c}(\tau)\right] ~,
	\label{PA_eqn27}
\end{align}
where $P_{\mathrm{c},0}$ is a special case as this can only happen for $\tau\leq T$.
The bunching probability $P_{\mathrm{b},n}$ takes a similar form as
\begin{align}
	P_{\mathrm{b},0} &= \left[\int_0^T d\tau \> \frac{T-\tau}{T}P_\mathrm{b}(\tau)\right]\nonumber \\*
	P_{\mathrm{b},n>0} &=
	\left[\int_{(n-1)T}^{nT} d\tau \> \frac{\tau-(n-1)T}{T}P_\mathrm{b}(\tau)\right.\nonumber \\*
	& \qquad \left.+\int_{nT}^{(n+1)T} d\tau \> \frac{(n+1)T-\tau}{T}P_\mathrm{b}(\tau)\right] ~.
	\label{PA_eqn28}
\end{align}

\subsection{Measurement probabilities}

Let us now consider how these fundamental probabilities translate to actual measurement outcomes for our different detector types.
At this point it is opportune to introduce the possibility of photon loss and limited detector efficiency, which can both be bundled into a single loss rate $\gamma$~\cite{Lyons-2018}.

\subsubsection{Bucket detectors}
For bucket detectors, when we have a coincidence we expect both detectors to click, provided neither photon is lost.
We also get a rough estimate of the time delay, up to the coarseness due to time-binning.
The probability of two clicks with $n$ time bin separation is given by $P_{2,n}^\mathrm{B}$.
If exactly one photon is lost, we will get a single click from one of the detectors, regardless of whether or not our photons bunched.
In the event of bunching, we will only ever get at most one click, as a single detector is not able to discriminate between one and two arriving photons.
The time bin information for a single click is irrelevant without knowing when the other photon arrived.
The probability of a single click (whether from loss or bunching) is $P_1^\mathrm{B}$.
Neither detector will click if both photons are lost, which happens with probability $P_0^\mathrm{B}$. 
We therefore get the following click probabilities:
\begin{align}
	P_{2,n}^\mathrm{B} &= (1-\gamma)^2 P_{\mathrm{c},n} ~,
	\label{PA_eqn29}\\
	P_1^\mathrm{B} &= 2\gamma(1-\gamma) P_{\mathrm{c},\mathrm{tot}}+(1-\gamma^2)P_{\mathrm{b},\mathrm{tot}} ~,
	\label{PA_eqn30}\\
	P_0^\mathrm{B} &= \gamma^2 ~,
	\label{PA_eqn31}
\end{align}
where as discussed the difference between time bins only applies to $P_{2,n}^\mathrm{B}$.

\subsubsection{Number-resolving detectors}

For number resolving detectors, we now have four types of potential outcomes to consider.
Barring photon loss, the probabililty of two clicks from a coincidence, $P_{2\mathrm{c},n}^{\mathrm{NR}}$, remains unchanged, but we now also have a probability for two clicks when the photons bunch, $P_{2\mathrm{b},n}^{\mathrm{NR}}$. The time bin information is relevant for both of these two-click outcomes and $n$ denotes the time bin separation.
If only one detector clicks, we have definitively lost exactly one photon, which occurs with probability $P_1^{\mathrm{NR}}$. Finally, probability of zero clicks remain unchanged and is now denoted by $P_0^{\mathrm{NR}}$. The full set of outcomes is therefore given by: 
\begin{align}
P_{2\mathrm{c},n}^{\mathrm{NR}} &= (1-\gamma)^2 P_{\mathrm{c},n} ~,\label{PA_eqn32}\\
	P_{2\mathrm{b},n}^{\mathrm{NR}} &= (1-\gamma)^2 P_{\mathrm{b},n} ~,\label{PA_eqn33}\\
	P_1^{\mathrm{NR}} &= 2\gamma(1-\gamma) ~,\label{PA_eqn34}\\
	P_0^{\mathrm{NR}} &= \gamma^2 ~.
	\label{PA_eqn35}
\end{align}

\section{Parameter estimation}\label{sec_para_est}

To address the question of how well we can extract $\delta$, and thus $\delta_s$, given a record of measurement data in the form described above, we turn to the theory of parameter estimation theory. 
The following section introduces the relevant concepts, at first in general terms and then applied to our specific scenario.

\subsection{Probability distributions}
The \ac{CRB} can be used to bound the variance of an unbiased estimator \( \estimator{\vec{\theta}} \) of unknown parameter(s) \( \vec{\theta} \) as~\cite[Chap.~3]{Kay-1993}
\begin{equation}
	\cov (\estimator{\vec{\theta}}) \geq \frac{1}{N}\vec{F}^{-1}(\vec{\theta}) ~,
	\label{CR_eqn1}
\end{equation}
where \( F(\vec{\theta}) \) is the \ac{FIM} and $N$ the number of repetitions of the experiment.
For a probability distribution \( P(m|\vec{\theta}) \) with outcomes \( m \in \mathcal{M} \) the \ac{FIM} is defined with elements
\begin{equation}
	[\vec{F}(\vec{\theta})]_{i,j} = \sum\limits_{m\in M} \frac{1}{P(m|\vec{\theta})} \frac{\partial}{\partial\theta_i} P(m|\vec{\theta}) \frac{\partial}{\partial\theta_j} P(m|\vec{\theta}) ~.
	\label{FI_eqn1}
\end{equation}
The \ac{MLE} is an estimator which is, in general, asymptotically consistent and efficient, meaning that in the limit \( N \to \infty \) it is both unbiased and its variance saturates the \ac{CRB}~\cite[Chap.~7]{Kay-1993}.

Each diagonal term of the \ac{FIM} $[\vec{F}(\vec{\theta})]_{i,i}$ corresponds to the single-parameter \ac{CFI} for the $i^{\mathrm{th}}$ element of $\vec{\theta}$, which we denote $F_{\theta_i}$ as shorthand.
This quantifies the amount of information about $\theta_i$ that is gained from an average individual measurement.
The off-diagonal terms represent covariance between the parameters, from which the correlations can be obtained.
These correlations can give rise to indeterminate \acp{FIM} for which the parameters \( \{ \theta_i \} \) cannot all be estimated simultaneously, while a subset of the parameters may still be estimable if the remaining parameters are known (e.g.\ through a calibration stage).

Our \ac{HOM} approach is parameterised by the set \( \theta_i \in \{ \delta,\alpha,\sigma,\gamma \} \). In the following we will be concerned with evaluating scalar and vector bounds for the different probability distributions (detector configurations) discussed in Sec.~\ref{sec_prob_analysis}.
Estimating \( \delta \) alone will be our primary focus, however we will also look at the potential for estimating calibration parameters independently of and in parallel with \( \delta \).
For the latter (multi-parameter) setting the rank of the \ac{FIM} is of particular value, identifying the number of independent parameters which can be estimated.

The traditional binary outcome \ac{HOM} protocol is a simple case where we can see that---operating at a fixed point in the \ac{HOM} dip---the coincidence rate can be modified by either changing \( \alpha \) or \( \delta \).
Hence, it is not possible to estimate \( \delta \) without knowing \( \alpha \), and this gives rise to the need for a calibration stage~\cite{Lyons-2018}.
We discuss this in more detail later and in Appendix~\ref{no_TR_FIMs}.

\subsection{Quantum states}

The statistics of different detectors can be considered as different POVMs acting on the same quantum state%
\footnote{Although we applied loss and time binning to the probabilities Eqs.~\eqref{PA_eqn24} and~\eqref{PA_eqn36}, these could equally be modelled by different POVMs.}.
The \ac{QFI} is an upperbound to the \ac{CFI} for quantum systems which depends only on the quantum state and so independent of the measurement used.
In the context of the \ac{CRB} it gives rise to the \ac{QCRB}~\cite{Braunstein-1994,Paris-2009,Toth-2014}
\begin{equation}
	\mathrm{var}(\tilde{\theta}) \geq \frac{1}{N F_\theta} \geq \frac{1}{N H_\theta} ~,
	\label{ID_eqn15}
\end{equation}
where we give only the single-parameter bound and \( H_{\theta} \) is the single-parameter \ac{QFI}.
For the purposes of parameters encoded in quantum states, it is now the \ac{QCRB} that sets the ultimate limit on the precision of an estimator.
The single-parameter \ac{QCRB} corresponds to the precision we would obtain from an optimal measurement~\cite{Braunstein-1994,Paris-2009}.
For a pure state, the single-parameter \ac{QFI} has the relatively simple form~\cite{Paris-2009}:
\begin{equation}
	H_{\theta} = 4[\braket{\partial_\theta\psi|\partial_\theta\psi}+(\braket{\partial_\theta\psi|\psi})^2] ~.
\label{ID_eqn2}
\end{equation}

\section{Benchmarks}\label{sec_ideal_detectors}

\subsection{Quantum Fisher information}
The parameter of particular interest to us is the time delay $\delta$. 
In order to establish an upper bound on the observable information of $\delta$, we now derive the \ac{QFI} for $\delta$ for our biphoton state.
Since the \ac{QFI} is invariant under parameter-independent unitary transformations~\cite{Toth-2014}, we are free to return to our biphoton state in the form given in Eq.~\eqref{PA_eqn18}, after it has picked up the time delay $\delta$ but before interaction at the beamsplitter.

By writing the biphoton spectral amplitude as univariate (as is customarily done for down-converted photon pairs) we have implicitly evaluated a delta function in the bivariate biphoton spectral amplitude,
which requires some additional care with regards to normalising the state.
To calculate the QFI, we thus introduce the quantisation volume $V$ and `renormalise' both the Dirac delta function such that $\delta_\mathrm{D}(0)=V$, and 
the joint spectral amplitude by the replacement $\phi(\omega)\to\phi(\omega)/\sqrt{V}$~\cite[Chap.~4,~6]{Loudon-2000}
Our normalised state now takes the form
\begin{align}
	\ket{\psi} &= \frac{1}{\sqrt{V}}\int d\omega \> \phi(\omega) e^{i\omega\delta} \nonumber \\*
	&  \times (\sqrt{\alpha}\>\hat{a}_i^\dagger(\omega)+\sqrt{1-\alpha}\>\hat{b}_i^\dagger(\omega))\hat{a}_s^\dagger(\omegap-\omega)\ket{0} ~.
	\label{ID_eqn8}
\end{align}

Equation~\eqref{ID_eqn2} can be used to calculate the \ac{QFI} from Eq.~\eqref{ID_eqn8} through 
\begin{equation}
	\braket{\partial_\delta\psi|\psi} 
	= -i \int d\omega \> \phi(\omega)^2 \omega 
	= -\frac{i\omegap}{2} ~,
	\label{ID_eqn3}
\end{equation}
and
\begin{equation}
	\braket{\partial_\delta\psi|\partial_\delta\psi} = \int d\omega \> \phi(\omega)^2 \omega^2 
	= \sigma^2 + \frac{\omegap^2}{4} ~,
	\label{ID_eqn4}
\end{equation}
full derivations for which are given in Appendix~\ref{qfiCalcs}.
We can then combine Eqs.~\eqref{ID_eqn2},~\eqref{ID_eqn3} and~\eqref{ID_eqn4} to obtain
\begin{equation}
	H_\delta=4\sigma^2 ~.
	\label{ID_eqn5}
\end{equation}

Our expression for the QFI agrees with that obtained in Ref.~\cite{Chen-2019} for frequency-entangled input photons when taking the limit of vanishing frequency detuning. We note that Eq.~\eqref{ID_eqn5} has no dependence on $\alpha$. 
Therefore, any optimal measurement protocol should be unaffected by the relative indistinguishability of the two photons.
Additionally, there is no dependence on $\delta$ itself: the maximum information should be obtainable regardless of the specific size of the delay.
These are both in contrast to our HOM protocols, where the \ac{CFI} depends strongly on both $\alpha$ and $\delta$, as we shall see in the following.

\subsection{Time of flight protocol (no-HOM)}
An optimal measurement scheme, which maximises the \ac{CFI} is, perhaps unsurprisingly, trivially obvious: we use time-resolving detectors with infinite precision.
No beamsplitter appears in this protocol, we eliminate the HOM effect entirely and simply place detectors at the ends of the paths for both photons.
This has the additional benefit that we never lose track of which photon is which, and can now even distinguish between positive and negative $\delta$.

We start by outlining this protocol and analysing its performance.
Using the same pre-beamsplitter state given in Eq.~\eqref{PA_eqn99} we take the set of \acp{POVM} $\Pi^{\mathrm{NH}}(\tau)$ (defined in Appendix~\ref{POVMs}) for which the probability {$P^\mathrm{NH}(\tau)=\bra{\psi}\hat{\Pi}^{\mathrm{NH}}(\tau)\ket{\psi}$} that our photons arrive at the detectors with temporal separation $\tau$ is, with appropriate normalisation,
\begin{equation}
	P^{\mathrm{NH}}(\tau) = \sqrt{\frac{2}{\pi}} \> \sigma \> e^{-2\sigma^2(\delta-\tau)^2}~.
	\label{ID_eqn7}
\end{equation}
The full derivation for this expression is detailed in Appendix~\ref{POVMs}.
Sampling directly from this distribution the \ac{CFI} matches the \ac{QFI} of \( 4\sigma^2 \).

We now apply the binning procedure as described in Sec.~\ref{sec_prob_analysis} to find the probability that our photons arrive $n$ time bins apart.
Now that as we can distinguish between positive and negative $\delta$ (as we know which arm each detected photon travelled along), it is important to distinguish positive and negative $n$. Therefore $\tau$ here is not simply a magnitude, as previously, but can itself be positive or negative.
The probability that the photon passing through the sample arrives $n$ time bins after the other photon is therefore
\begin{align}
	P^\mathrm{NH}_n&=\int_{(n-1)T}^{nT} d\tau \> \frac{\tau-(n-1)T}{T}P^\mathrm{NH}(\tau) \nonumber \\* 
	& \qquad +\int_{n\tau}^{(n+1)T} d\tau \> \frac{(n+1)T-\tau}{T}P^\mathrm{NH}(\tau) ~.
	\label{ID_eqn11}
\end{align}
We insert the photon loss rate $\gamma$ and obtain the measurement probabilities
\begin{align}
	P_{2,n}^\mathrm{NH}&=(1-\gamma)^2 P^\mathrm{NH}_n ~,
	\label{ID_eqn12}\\
	P_1^\mathrm{NH}&=2\gamma(1-\gamma) ~,
	\label{ID_eqn13}\\
	P_0^\mathrm{NH}&=\gamma^2 ~,
	\label{ID_eqn14}
\end{align}
for two clicks (with $n$ bin separation), one click, or zero clicks, respectively.

We will refer to this protocol as `no-HOM' and use it as a benchmark of the raw time resolution of the detectors alone. We shall see that whilst it performs much worse than \ac{HOM} protocols at low and moderate detector time resolution, it can beat the \ac{HOM} approach at a sufficiently high resolution. Note that this is similar to the time-of-flight protocols of Ref.~\cite{Pellegrini-2000,tobin-2017,Korzh-2020}, where direct timing information is used to construct an image. The typical resolutions involved in such a set-up are $10$-$30 \text{ ps}$, although Ref.~\cite{Korzh-2020} reports a temporal resolution of $\simeq 2.6 \text{ ps}$.
Taking the limit of infinite temporal resolution for the no-HOM approach yields the optimal protocol, and with $\gamma=0$ we find $F_\delta \to H_\delta$ as $T\to0$ for all $\delta$ and $\alpha$.

\subsection{Loss-adjusted quantum Fisher information}

For $\gamma\neq0$ we note that information can only be obtained when neither photon is lost, which happens with probability $(1-\gamma)^2$.
Allowing only for measurements to happen when both photons are detected, the effective \ac{QFI} will be reduced by the probability of neither photon being lost
\begin{equation}
	H^{\mathrm{2ph}}_\delta(\gamma) = (1-\gamma)^2 H_\delta = 4 \sigma^2 (1-\gamma)^2.
	\label{ID_eqn16}
\end{equation}
This ``two-photon conditioned \ac{QFI}'' is distinct from the actual \ac{QFI} of the loss-affected mixed state:
the \ac{QFI} for the mixed state subjected to loss will be larger, as when a single photon is lost, the state of the other photon still possesses some parameter-dependence.
However, that larger information is not readily accessible without additional information---such as the time at which the photons were initially generated---or resources in the measurement%
\footnote{%
Similar subtleties are observed in phase estimation where additional resources or knowledge can be required to realise \acp{QFI}~\cite{Monras-2006,Jarzyna-2012}.
}.
As we are principally interested in the relative delay between the two photon arrival times rather than the absolute length of either path we favour using $\Htwoph$ as a loss-adjusted point of comparison.

\section{Results}\label{sec_results}

In the following, it will be convenient to express our temporal parameters $\delta$ and $T$ in units of $1/\sigma$, i.e.~in units of the inverse of the photons' spectral width which is equal to the width of the \ac{HOM} dip.
This yields the \ac{CFI} $F_\delta$ in units of $\sigma^2$, though note that we will usually rescale this as a fraction of $\Htwoph$, the two-photon conditioned \ac{QFI}.
We refer to this rescaled quantity, $F_\delta/\Htwoph$, as the relative information $\Irel$.

\begin{figure*}
\centering
\includegraphics[width=\linewidth]{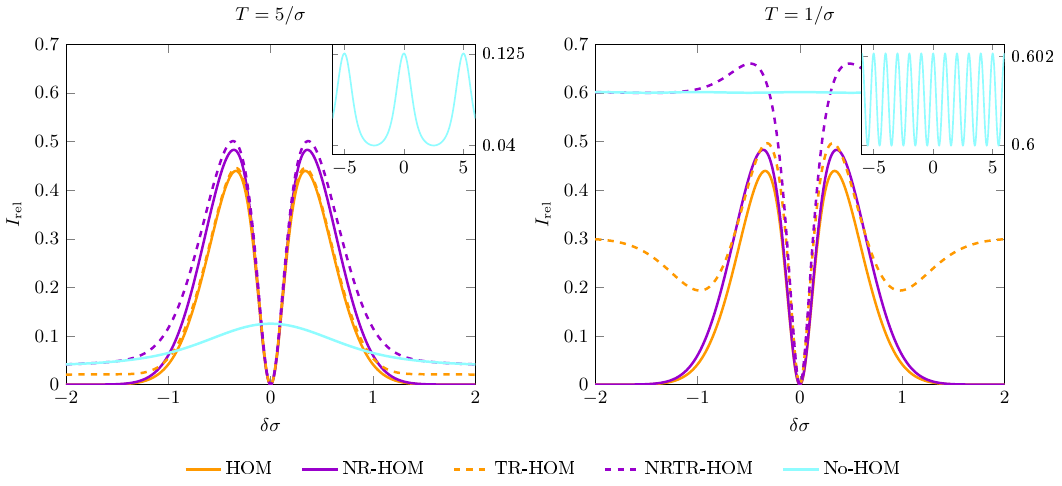}
\caption{Comparison of the relative information $\Irel=F_\delta/\Htwoph$, the \acl{CFI} as a fraction of the two-photon conditioned \acl{QFI}, for our protocol configurations as we vary the delay $\delta$. We contrast the basic HOM protocol with bucket detectors against versions with number-resolving  (NR-HOM) and time-resolving (TR-HOM) detectors, as well both enhancements simultaneously (NRTR-HOM). Finally, no-HOM is a time of flight protocol that solely measures arrival times.
We set $\alpha = 0.9$ and $\gamma = 0.4$, and the time bin widths $T = 5/\sigma$ on the left, and $T = 1/\sigma$ on the right.
All HOM protocols feature peaks in the information equally spaced around $\delta=0$. The no-HOM protocol displays (non-sinusoidal) oscillation in the relative information with period equal to $T$.
The amplitude of this oscillation reduces with low $T$, and can be seen in more detail in the insets.
The right plot shows the benefits of reduced $T$ (better resolution) by the increased information.}
\label{results_fig1}
\end{figure*}

Only for detectors without time-resolution are we able to obtain relatively neat closed-form expressions for the \ac{CFI}.
We obtain the single parameter \acp{CFI} for $\delta$ (as well as multi-parameter \ac{CFI} matrices) in Appendix~\ref{no_TR_FIMs}.
The time-delay \acp{CFI} are given by Eqns.~\eqref{no_TR_FIMs_eqn5} and \eqref{no_TR_FIMs_eqn6}.
For the cases with time-resolution, the \acp{CFI} are unfortunately less amenable for analytical inspection, but can be straightforwardly evaluated numerically. 

We compare four different configurations of the \ac{HOM} protocol: First, the standard HOM protocol (such as that in Ref.~\cite{Lyons-2018}) involving bucket detectors and no time-resolution; this will be referred to as the \ac{HOM} protocol without qualifiers. We assign the label NR-HOM to a protocol that is enhanced with number-resolving detectors. Further, we consider two variants without and with time-resolution, referred to as TR-HOM and NRTR-HOM, respectively. The no-HOM protocol outlined in Sec.~\ref{sec_ideal_detectors} will serve as a benchmark throughout.

\subsection{Optimal delay}

In Fig.~\ref{results_fig1} we compare $\Irel$ for our different protocols as a function of $\delta$. 
All \ac{HOM} protocols with visibilty $\alpha<1$ feature two characteristic peaks that are symmetric about $\delta=0$ with peak positions that 
differ slightly between the different protocols.
At small values of $\delta$ there is a dip in $\Irel$, dropping all the way to zero at $\delta=0$.
Adjusting $\delta_a$ allows us to tune $\delta$ such that it falls into a favourable local environment near the maximal \ac{CFI}, enabling operation close the optimal point.
In practical scenarios such as those in Refs.~\cite{Lyons-2018,Lyons-2018_2,Giovannini-2015}, $\delta_s$ is known to be sufficiently small that we can confidently tune ourselves near this optimal point without requiring significant additional resources. We shall likewise make this assumption going forward.
The no-HOM protocol is also shown and indicates the background information coming purely from the time bin information of our measurements (for the case of detectors with time resolution).
No-HOM does not suffer the same information dip at low $\delta$ that appears in all HOM protocols.

There is a slight (non-sinusoidal) oscillation in the no-HOM information, with period equal to the time bin width.
This emerges as it is optimal to have $\delta=nT$, for some integer $n$.
This reduces the variance in the measured bin difference, and the most likely result will be a difference of $n$ bins.
By comparison, for $\delta=(n+\frac{1}{2})T$, we will now see results primarily split between $n$ and $n+1$ bins, leading to a reduction in information obtained; this effect is substantial for $T=5/\sigma$.
As $T$ shrinks, however, the no-HOM background information increases
and the amplitude of these oscillations rapidly shrinks, becoming mostly negligible for $T \lesssim 1/\sigma$.

\begin{figure}[htbp]
\centering
\includegraphics[width=\linewidth]{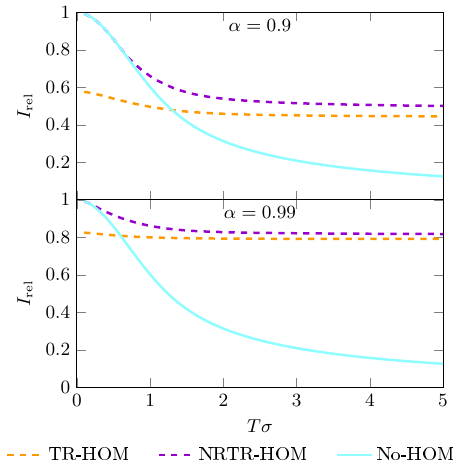}
\caption{The relative information $\Irel$ for time-resolving protocols as we vary time bin size $T$.
We set $\gamma = 0.4$, and $\delta$ is chosen to maximise the information; with $\alpha = 0.9$ (as in Fig.~\ref{results_fig1}) in the top panel and $\alpha = 0.99$ in the bottom panel.
The no-HOM protocol performs poorly at low time-resolution, but around $T=1/\sigma$ overtakes TR-HOM and achieves parity with NRTR-HOM.
As $T \to 0$, no-HOM and NRTR-HOM approach the QFI limit.
Increased visibility extends the range across which a HOM-based protocol is advantageous.}
\label{results_fig2}
\end{figure}

At large delay, $\Irel$ for NRTR-HOM tends towards the no-HOM information, as we no longer get useful information from coincidence and bunching rates as the temporal modes become orthogonal.
For TR-HOM, however, it instead tends towards half the no-HOM information, as half the time our photons will bunch and we can only detect the arrival of one of those photons, a measurement which carries no useful delay information.

\subsection{Time-resolution}
As we increase temporal resolution, time-resolving protocols gain an increasing advantage over their less advanced counterparts.
In Fig.~\ref{results_fig1}, looking at the maximal information, no-HOM performs worse than all HOM protocol configurations with $T=5/\sigma$. However, with $T=1/\sigma$ no-HOM only performs narrowly worse than NRTR-HOM but now beats our other HOM protocols.
We can see more clearly in Fig.~\ref{results_fig2} how the maximal information varies with time resolution.
Given NRTR-HOM contains the same timing information of no-HOM with additional information coming from the bunching rate, it will never perform worse.
However, as $T\to0$, timing information dominates and HOM information becomes irrelevant.
Thus the two protocols achieve parity in maximal information, and both tend towards the \ac{QFI}.

\begin{figure}[htbp]
\centering
\includegraphics[width=\linewidth]{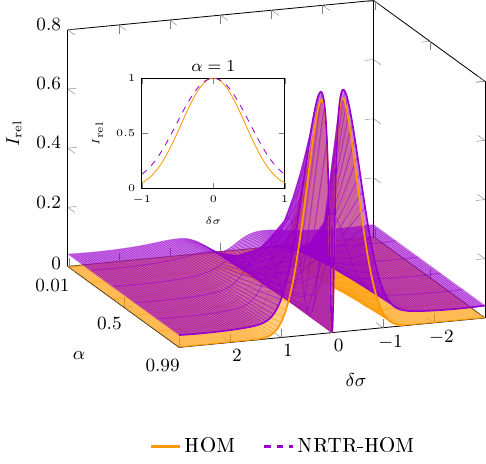}
\caption{The relative information $\Irel$ as we vary $\delta$ and $\alpha$ for the two extreme HOM protocols: standard HOM (lower surface) vs NRTR-HOM (upper surface).
The two characteristic peaks grow and move inwards at higher visibility; merging to a single peak at the origin that saturates the two-photon conditioned \ac{QFI} for perfect visibility $\alpha =1$ as shown in the inset.
At large $\delta$,  background time bin information manifests itself in the NRTR-HOM surface not falling to zero. 
Other parameters are $\gamma = 0.4$ and  $T = 5/\sigma$.
}
\label{results_fig8}
\end{figure}

Without NR capabilities there is a transition between TR bucket detectors being best placed in the TR-HOM or no-HOM configurations.
In Fig.~\ref{results_fig2} this is around $T \sim 1/\sigma$, at higher time-resolutions than which the no-HOM configuration is preferable.
TR-HOM falls short in this regime, as time bin information cannot be obtained from bunching events.
This transition is affected by the visibility $\alpha$ (which limits the effectiveness of the HOM but not no-HOM protocols) with increased $\alpha$ allowing TR-HOM to perform competitively at higher time-resolutions.

It is worth recalling that conventional photon detectors have time-resolving abilities far worse than $1/ \sigma$ of typical \ac{SPDC} photons:
whilst time-resolution was not used in Ref.~\cite{Lyons-2018}, the estimated detector timing jitter on the order of nanoseconds results in $T \sigma > \mathcal{O} (10^2)$.
Consequently, there would have been only negligible timing background information in those experiments, and \ac{HOM} effect information overwhelmingly dominated. 

\begin{figure*}
\centering
\includegraphics[width=\linewidth]{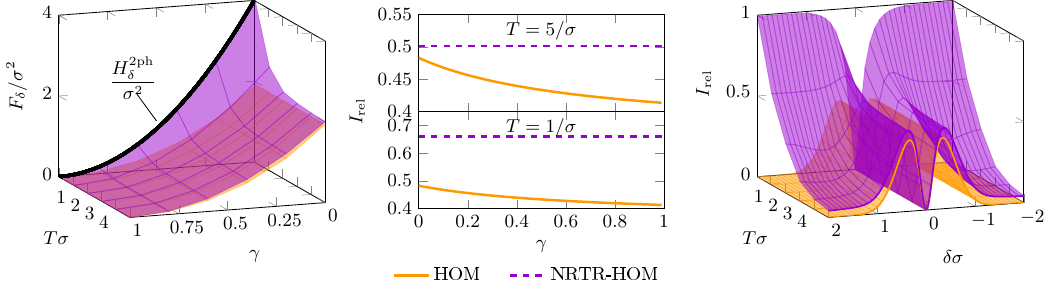}
\caption{Left: $F_\delta$ as a function of $\gamma$ and $T$. The standard HOM case (lower surface) is independent of $T$ and thus only depends on $\gamma$. 
Here, $\alpha = 0.9$ and $\delta$ is chosen to maximise information at each point.
Both protocols naturally perform better at lower $\gamma$, and NRTR-HOM (upper surface) approaches $\Htwoph$ as $T\to0$.
Centre: $\Irel$ as a function of $\gamma$. $\delta$ is chosen to maximise information at each point.
For NRTR-HOM, the relative information remains constant as $\gamma$ is varied, but for standard HOM the information decreases at higher $\gamma$ (see text).
$T=5/\sigma$ for the top plot, $T=1/\sigma$ for the bottom plot, $\alpha=0.9$ is used in both.
Right: the relative information $\Irel$, taking $\gamma=0.4$, and varying $\delta$ and $T$.
Standard HOM (lower surface) is again independent of T.
At vanishingly small $T$, NRTR-HOM (upper surface) achieves the QFI limit for sufficiently large $\delta$.}
\label{results_fig9}
\end{figure*}

\subsection{Visibility \& photon loss}

In the following we focus on the two most extreme variants among our \ac{HOM} protocols: the `all bells and whistles' case of NRTR-HOM contrasted against the conventional \ac{HOM} scenario.
In Fig.~\ref{results_fig8} we can see how the information for our NRTR-HOM protocol varies with both $\delta$ and $\alpha$.
The two peaks rise rapidly as visibility is increased, moving inwards and converging to a single peak at the origin at the limit $\alpha=1$, achieving the \ac{QFI} limit at this point.
At large delays information tends to the lower background level, that is obtained solely from arrival times and is thus independent of $\alpha$.
Our standard HOM protocol follows a similar shape though performs slightly worse.
At large delays its information falls to zero, without the background information coming from simple arrival time data.

The central plot of Fig.~\ref{results_fig9} shows how the relative information $\Irel$ varies with $\gamma$.
For NRTR-HOM, the \ac{CFI} scales linearly with $(1-\gamma)^2$, the same dependence as in the \ac{QFI}.
Thus $\Irel$ is constant with $\gamma$.
Standard HOM performs even worse at increased photon loss rates as compared to the two-photon conditioned \ac{QFI}, due to one-click measurements not only dominating over two-clicks events but also becoming less likely to indicate bunching. This reduces the available information that can be gleaned and renders most of the measurement outcomes useless.

\subsection{Approaching the QFI}
In the left plot of Fig.~\ref{results_fig9}, we see that by letting the time bin width tend to $0$ we can, with NRTR-HOM, approach the limit of the two-photon conditioned \ac{QFI}.
This then achieves the full \ac{QFI} of $4\sigma^2$ with the additional realisation of $\gamma=0$.
The right plot of Fig.~\ref{results_fig9} shows that we achieve this for all sufficiently large $\delta$, though interestingly the dip in $\Irel$ remains around $\delta=0$.
As previously discussed, this is not much of a constraint, however, given the ability to adjust $\delta_a$.

\subsection{Multi-parameter estimation}
As discussed in Sec.~\ref{sec_para_est}, we may want to estimate multiple parameters as a means of eliminating the need for a separate calibration stage.
The most critical and tricky calibration parameter is $\alpha$---loss and spectral width are easier to determine, and at least the latter is less variable.
For this reason we begin with the case of wishing to estimate $\delta$ and $\alpha$ simultaneously.

\begin{figure}[htbp]
\centering
\includegraphics[width=\linewidth]{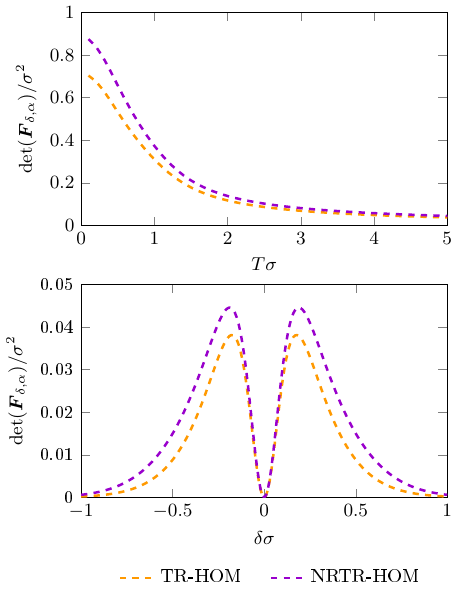}
\caption{The determinant of the \acl{FIM} for parameters $\delta$ and $\alpha$ for $\alpha = 0.9$ and $\gamma = 0.4$.
Top: for fixed  $\delta = 0.2/\sigma$ and varying time bin size $T$.
At higher time-resolution (smaller $T$) the determinant increases, and the benefit of number-resolving detectors become more pronounced.
Bottom: for fixed $T = 5/\sigma$ and varying $\delta$.
The two peaks in the determinant represent the optimal points for estimating $\delta$ and $\alpha$ simultaneously.}
\label{results_fig7}
\end{figure}

Delay and visibility become increasingly independent at higher time-resolution, as we are better able to estimate $\delta$ independent of any coincidence/bunching information.
By constructing the \ac{FIM} for these two parameters, we see in Fig.~\ref{results_fig7} how this effect manifests as the determinant increases as time bin width shrinks.
This is analagous to an increase in the single-parameter \ac{CFI}, indicating increased information and thus making estimation more efficient.
For a fixed time bin width, if we instead vary $\delta$ we see a similar shape to that in Fig.~\ref{results_fig1}.

The optimal points for estimating $\delta$ and $\alpha$ simultaneously are given by the peaks in the determinant.
The requirement that we now estimate $\alpha$ alongside $\delta$ causes the peaks to shift inwards as compared to those in Fig.~\ref{results_fig1}, as $\alpha$ itself is best estimated close to the origin (see Appendix~\ref{other_paras} for further discussion). 

Further investigating the eigenvalues can give a clearer idea of what the independently estimable parameters may be and their attainable precision~\cite{Ragy-2016,Bisketzi-2019}, and even reveal `sloppiness': how many effective parameters are required to capture a model's behaviour~\cite{Davidson-2020}.
Our focus has been on estimating \( \delta \) and so it may be beneficial to cast the other parameters as nuisance parameters~\cite{Lehmann-1998,Suzuki-2020}, however there are cases where tracking a parameter such as visibility is of independent interest~\cite{Cimini-2019,Cimini-2019_2}.

\section{Discussion and conclusion}\label{sec_discussion}

Our analysis shows how Hong-Ou-Mandel metrology can be boosted by the introduction of more sophisticated detector hardware. In particular, a switch from bucket detectors (i.e.\ detectors that are incapable of resolving more than one photon simultaneously or with a short time gap due to dead time)  to number-resolving detectors yields significant improvements regardless of the detector time-resolving capabilities, an immediate boost in Fisher information of \SI{9.9}{\percent} at our benchmark of $\alpha=0.9$, $\gamma=0.4$. This increase relative to bucket performance is even further pronounced at lower visibility and higher photon loss.
Number-resolving detectors have historically suffered additional drawbacks, such as extremely low operating temperatures making them costly to implement~\cite{Marsili-2009,Eisaman-2011}, however, recent developments~\cite{Young-2019,Zhu-2019,Kardynal-2008} suggest such detectors may soon become more viable.

Unlocking the increase in accessible information offered by time-resolving detectors may present a more significant challenge. Taking \citet{Lyons-2018} as an example: here the \SI{808}{\nm} downconverted photons, together with a \SI{10}{\nm} bandpass filter, lead to a dip of spatial width \SI{\approx 65}{\um}. This corresponds to a temporal width of \SI{\approx 0.2}{\ps} and spectral width $\sigma\approx \SI{4.6}{\ps^{-1}}$.
To see a noticeable increase in precision would require detector resolution approaching the order of \SI{1}{\ps}, leading to an improvement of around \SI{1.3}{\percent}, or \SI{14}{\percent} if  the detectors are also number-resolving.
This is as-of-yet beyond the precision offered by conventional modern detectors though some recent work is nearing this level of precision. Work by \citet{Zadeh-2018,Korzh-2020} show \acfp{SNSPD} offering sub-\SI{10}{\ps} resolution while retaining high detector efficiency. Whilst the precision increase from these particular detectors would still be modest, this nonetheless already opens up the possibility of a multi-parameter estimation scheme in order to eliminate the need for determining the width, loss, and most importantly visibility calibration parameters separately.

It would take a detector time resolution as high as \SI{0.2}{\ps} (or, more generally, on the order of the \ac{HOM} dip width) for no-HOM approach,
to acquire competitive performance:
no-HOM would then perform \SI{36}{\percent} better than HOM, and also beat TR-HOM,
but still fall short of NRTR-HOM which boast a \SI{50}{\percent} increase over HOM. 
If resolution were further doubled, no-HOM would perform nearly on par with time- and number-resolving \ac{HOM} which features an increase of \SI{95}{\percent} over basic \ac{HOM}.
Therefore, detectors with such high time resolution would obviate the need for a \ac{HOM} protocol altogether: in this case almost all information can be obtained by directly measuring arrival times, and extra information gained from coincidence/bunching results becomes negligible, and this would moot the advantage of number-resolving detectors, which are unnecessary in a no-HOM setup. However, this level of temporal resolution (combined with the desirable high efficiency and low dark counts) is unlikely to become available in the near future, if ever, so \ac{HOM} interferometry will likely remain the uncontested approach towards estimating phase-non-sensitive group velocity delays.

At this point, it is worth discussing a few simplifying assumptions which we have made in our analysis: first, we have assumed exactly one input photon pair per coincidence matching time window.
In reality, \ac{SPDC} does not generate photons at a constant rate and could, at least in CW mode, generate two pairs in rapid succession, or none at all for a given window, with the latter case not affecting our benchmarking conclusions, but bearing on the rate at which information is accumulated.
Next, for our bucket detectors we have assumed that these have a dead time exceeding any possible delay between the two photons, i.e. such a detector could only ever register at most one photon from a given pair.
Finally, we have omitted any consideration of detector dark counts: instances where a detector randomly clicks without a photon arrival.
The prevalence of dark counts imposes limits on the rate at which we want to generate photons: too low a rate and dark counts will dominate, too high and we risk falling foul of dead times or associating photons from different pairs within the same coincidence window.
These are all key things to consider in any experimental setup, but should have negligible impact on our benchmarking efforts to compare performance, provided an appropriate rate of photon generation is chosen.

As a final note, we shall compare the respective benefits of time- and number-resolution within the limits of available technology.
We will as previously assume typical values for the temporal dip width of \SI{0.2}{\ps}, corresponding spectral width $\sigma = \SI{4.6}{\ps^{-1}}$, and a visibility of $\alpha=0.9$.
For our time-resolving benchmark we take the \acf{SNSPD} detailed in \cite{Zadeh-2018}, a detector without number-resolving capabilities. This detector boasts an impressive \SI{86}{\percent} detection efficiency or, equivalently, a photon loss rate of $\gamma=0.14$\footnote{In reality, we would expect $\gamma$ to be higher due to other factors contributing to photon loss, however detector efficiency will dominate. It is worth again noting, however, that the performance of number-resolving detectors is less sensitive to loss, and any unaccounted for loss outwith   the detectors would have a more serious effect on the performance of the bucket detectors.}
\acp{SNSPD} can achieve sub-\SI{10}{\ps} resolution, and taking \SI{10}{\ps} resolution as a conservative estimate we obtain a maximal Fisher information $F_\delta=\SI{29.014}{\ps^{-2}}$. At the limit of multiphoton excitation, \acp{SNSPD} resolution can be pushed to sub-\SI{3}{\ps}. This corresponds to only a marginal increase in information, however, as at \SI{3}{\ps} we see $F_\delta=\SI{29.096}{\ps^{-2}}$.
For our number-resolving benchmark we shall consider the \acfp{TES} as described in \cite{Lita-2009}.
\acp{TES} have number-resolving capabilities but large enough timing jitter that we can consider them to have no useful time-resolution capabilities for the purposes of \ac{HOM} metrology. They are capable of achieving detector efficiency higher than \SI{95}{\percent}, so we take $\gamma=0.05$.
This detector can achieve a maximal $F_\delta=\SI{36.897}{\ps^{-2}}$.

In summary, while \ac{HOM} metrology has often been presented as a means to circumvent limitations in detector resolution, we are now in fact nearing the point where state-of-the-art detectors will possess sufficient resolution to augment the conventional \ac{HOM} protocol for resolving path length delays.
These gains will be gradual, as resolution improves. Whilst engineering a narrower or more structured \ac{HOM} dip can boost the performance of \ac{HOM} protocols, further extending their lead over direct time-of-flight measurements, in situations where a sufficiently narrow dip is not possible, the need for \ac{HOM} approach may be eliminated in the longer-term should the native detector resolution surpass the order of the dip width.
Likewise, we have shown that \ac{HOM} protocols can be significantly enhanced through the adoption of number-resolving detectors, an immedate improvement that could be implemented with current technology. Further, number-resolving detectors can allow \ac{HOM}-based protocols to remain superior even as timing resolution is improved.
Perhaps most importantly in the near future, our analysis highlights the possibility of simplified calibration-free \ac{HOM} metrology and points the way toward asymptotically approaching the precision of the \ac{QCRB}.

\begin{acknowledgments} The authors thank George Knee for discussions and feedback. This work was supported by UK EPSRC Grant EP/R030413/1. NW also wishes to acknowledge financial support from UK EPSRC Grant EP/R513222/1.
\end{acknowledgments}

\appendix

\onecolumngrid

\section{Arrival rates}\label{POVMs}

We have the following \ac{POVM} elements for coincidence and bunching events with a delay \( \tau \) expressed in terms of the average and difference in arrival times \( t \) and \( \tau \).
For coincidence events, considering only two-photon events, we have:
\begin{align}
	\hat{\Pi}_\mathrm{c}(\tau) = 2\pi\bigg[&\hat{c}_1^\dagger \left(t-\frac{\tau}{2}\right) \hat{c}_2^\dagger \left(t+\frac{\tau}{2}\right) \ket{0}\bra{0} \hat{c}_2 \left(t+\frac{\tau}{2}\right) \hat{c}_1\left(t-\frac{\tau}{2}\right) 
	+\hat{c}_1^\dagger \left(t-\frac{\tau}{2}\right) \hat{d}_2^\dagger \left(t+\frac{\tau}{2}\right) \ket{0}\bra{0} \hat{d}_2\left(t+\frac{\tau}{2}\right) \hat{c}_1 \left(t-\frac{\tau}{2}\right) \nonumber \\*
	+&\hat{d}_1^\dagger \left(t-\frac{\tau}{2}\right) \hat{c}_2^\dagger \left(t+\frac{\tau}{2}\right) \ket{0}\bra{0} \hat{c}_2\left(t+\frac{\tau}{2}\right) \hat{d}_1\left(t-\frac{\tau}{2}\right) 
	+\hat{c}_1^\dagger \left(t+\frac{\tau}{2}\right) \hat{c}_2^\dagger \left(t-\frac{\tau}{2}\right) \ket{0}\bra{0} \hat{c}_2 \left(t-\frac{\tau}{2}\right) \hat{c}_1\left(t+\frac{\tau}{2}\right) \nonumber \\*
	+&\hat{c}_1^\dagger \left(t+\frac{\tau}{2}\right) \hat{d}_2^\dagger \left(t-\frac{\tau}{2}\right) \ket{0}\bra{0} \hat{d}_2\left(t-\frac{\tau}{2}\right) \hat{c}_1 \left(t+\frac{\tau}{2}\right) 
+\hat{d}_1^\dagger \left(t+\frac{\tau}{2}\right) \hat{c}_2^\dagger \left(t-\frac{\tau}{2}\right) \ket{0}\bra{0} \hat{c}_2\left(t-\frac{\tau}{2}\right) \hat{d}_1\left(t+\frac{\tau}{2}\right)\bigg] ~,
\label{PA_eqn21}
\end{align}
whereas for bunching events, considering only two-photon events, we have:
\begin{align}
	\hat{\Pi}_\mathrm{b}(\tau) = &2\pi\sum\limits_{j=1}^2\bigg[\hat{c}_j^\dagger \left(t-\frac{\tau}{2}\right) \hat{c}_j^\dagger \left(t+\frac{\tau}{2}\right) \ket{0}\bra{0} \hat{c}_j \left(t+\frac{\tau}{2}\right) \hat{c}_j\left(t-\frac{\tau}{2}\right) 
	+\hat{c}_j^\dagger \left(t-\frac{\tau}{2}\right) \hat{d}_j^\dagger \left(t+\frac{\tau}{2}\right) \ket{0}\bra{0} \hat{d}_j\left(t+\frac{\tau}{2}\right) \hat{c}_j \left(t-\frac{\tau}{2}\right) \nonumber \\*
	&\mkern96mu+\hat{d}_j^\dagger \left(t-\frac{\tau}{2}\right) \hat{c}_j^\dagger \left(t+\frac{\tau}{2}\right) \ket{0}\bra{0} \hat{c}_j\left(t+\frac{\tau}{2}\right) \hat{d}_j\left(t-\frac{\tau}{2}\right)\bigg] ~.
\label{PA_eqn22}
\end{align}
The factor of \( 2\pi \) introduced in Eqs.~\eqref{PA_eqn21} and~\eqref{PA_eqn22} comes from accounting for the average arrival time \( t \).
Due to the monovariate spectral distribution \( \phi( \omega ) \) for the biphoton state the average arrival time \( t \) drops out of the overlaps calculated below, in order to account for all detection events this \( 2\pi \) factor is introduced.
The same probabilities can be derived with a bivariate spectral distribution \( \phi( \omega_1 , \omega_2 ) \) in which case the integral over \( t \) can be performed explicitly.

The positivity of \( \hat{\Pi}_{\mathrm{c}}(\tau) \) and \( \hat{\Pi}_{\mathrm{b}}(\tau) \) is straightforward to see as they are simply summations of projectors onto the orthogonal states.
The orthogonality follows from the only non-zero commutators
\begin{align}
	[\hat{c}_j(t),\hat{c}^{\dagger}_k(t')] &= \delta_{j,k} \delta_{\mathrm{D}}(t-t') , & [\hat{d}_j(t),\hat{d}^{\dagger}_k(t')] = \delta_{j,k} \delta_{\mathrm{D}}(t-t').
\end{align}
Strictly these \ac{POVM} elements must be accompanied by an element \( \mathbb{1} - \int d\tau (\hat{\Pi}_{\mathrm{b}}(\tau)+\hat{\Pi}_{\mathrm{c}}(\tau)) \) to form a complete \ac{POVM}, however for states of form Eq.~\eqref{PA_eqn20} this occurs with probability zero (Eqs.~\eqref{PA_eqn25} and~\eqref{PA_eqn26} sum to \( 1 \)) and so we omit this term from the following calculation.

\subsection{HOM coincidence probability}
In order to fully derive the fundamental coincidence rate we must evaluate ${P_\mathrm{c}(\tau)=\bra{\psi}\hat{\Pi}_\mathrm{c}(\tau)\ket{\psi}}$, with $\ket{\psi}$ as given in Eq.~\eqref{PA_eqn20} and $\hat{\Pi}_\mathrm{c}(\tau)$ the \acp{POVM} given in Eq.~\eqref{PA_eqn21}.
We also note the following commutation relations~\cite[Chap.~6]{Loudon-2000}:
\begin{equation}
	[\hat{c}_i(\omega),\hat{c}_j^\dagger(t)]
	= [\hat{d}_i(\omega),\hat{d}_j^\dagger(t)]
	= \frac{1}{\sqrt{2\pi}}\delta_{ij} e^{-i\omega t},
	\label{POVMs_eqn1}
\end{equation}
which follow from
\[
	[\hat{c}_i(\omega),\hat{c}_j^\dagger(t)] = \frac{1}{\sqrt{2\pi}}\delta_{ij}\int d\omega' e^{-i\omega't}[\hat{c}_i(\omega),\hat{c}_j^\dagger(\omega')]
	= \frac{1}{\sqrt{2\pi}}\delta_{ij}\int d\omega' e^{-i\omega't}\delta(\omega-\omega')
	= \frac{1}{\sqrt{2\pi}}\delta_{ij} e^{-i\omega t}.
\]

For convenience, we relabel $t\to t_0 +\tau/2$, and first evaluate
\begin{align}
	\bra{\psi}\hat{c}_1^\dagger(t_0)\hat{c}_2^\dagger(t_0+\tau)\ket{0} 
	=& \frac{1}{2}\bra{0}\int d\omega \> \phi(\omega) e^{-i\omega\delta} \sqrt{\alpha} 
	[-i\hat{c}_1(\omega)+\hat{c}_2(\omega)][\hat{c}_1(\omegap-\omega)-i\hat{c}_2(\omegap-\omega)] 
	\hat{c}_1^\dagger(t_0)\hat{c}_2^\dagger(t_0+\tau)\ket{0}\nonumber \\
	=& \frac{1}{2\pi}\frac{\sqrt{\alpha}}{2}\int d\omega \> \phi(\omega) e^{-i\omega\delta} 
	(-e^{-i\omega t_0}e^{-i(\omegap-\omega)(t_0+\tau)}+e^{-i(\omegap-\omega)t_0}e^{-i\omega(t_0+\tau)}) \nonumber \\
	=& \frac{\sqrt{\alpha}}{4\pi}e^{-i\omegap t}\int d\omega \> \phi(\omega) e^{-i\omega\delta}
	(e^{-i\omega\tau}-e^{-i\omegap\tau}e^{i\omega \tau}).
	\label{POVMs_eqn2}
\end{align}
Then, multiplying Eq.~\eqref{POVMs_eqn2} by its conjugate we obtain
\begin{align}
	\bra{\psi}\hat{c}_1^\dagger(t_0)\hat{c}_2^\dagger(t_0+\tau)\ket{0}\bra{0}\hat{c}_2(t_0+\tau)\hat{c}_1(t_0)\ket{\psi} 
	&= \frac{\alpha}{16\pi^2}\int d\omega_1 \> \int d\omega_2 \> \phi(\omega_1)\phi^*(\omega_2) e^{-i(\omega_1-\omega_2)\delta} \nonumber \\*
	&\mkern128mu\times (e^{-i\omega_1\tau}-e^{-i\omegap\tau}e^{i\omega_1 \tau})(e^{i\omega_2\tau}-e^{i\omegap\tau}e^{-i\omega_2 \tau})\nonumber \\
	&= \frac{\alpha\sigma}{\sqrt{32\pi^3}} \> e^{-2\sigma^2(\delta+\tau)^2}(e^{4\delta\sigma^2\tau}-1)^2
	\label{POVMs_eqn3}
\end{align}

Next, we evaluate
\begin{align}
	\bra{\psi}\hat{c}_1^\dagger(t_0)\hat{d}_2^\dagger(t_0+\tau)\ket{0} 
	=& \frac{1}{2}\bra{0}\int d\omega \> \phi(\omega) e^{-i\omega\delta} 
	(\sqrt{\alpha}[-i\hat{c}_1(\omega)]+\sqrt{1-\alpha}[\hat{d}_2(\omega)])[\hat{c}_1(\omegap-\omega)] 
	\hat{c}_1^\dagger(t_0)\hat{d}_2^\dagger(t_0+\tau)\ket{0}\nonumber \\
	=& \frac{\sqrt{1-\alpha}}{4\pi}\int d\omega \> \phi(\omega) e^{-i\omega\delta} 
	e^{-i(\omegap-\omega)t_0}e^{-i\omega(t_0+\tau)}\nonumber \\
	=& \frac{\sqrt{1-\alpha}}{4\pi}\int d\omega \> \phi(\omega) e^{-i\omega\delta} 
	e^{-i\omegap t_0}e^{-i\omega\tau}.
	\label{POVMs_eqn4}
\end{align}
Again, multiplying by its conjugate we find
\begin{align}
	\bra{\psi}\hat{c}_1^\dagger(t_0)\hat{d}_2^\dagger(t_0+\tau)\ket{0}\bra{0}\hat{d}_2(t_0+\tau)\hat{c}_1(t_0)\ket{\psi} 
	=& \frac{1-\alpha}{16\pi^2}\int d\omega_1 \> \int d\omega_2 \> \phi(\omega_1)\phi^*(\omega_2) e^{-i(\omega_1-\omega_2)\delta} 
	(e^{-i\omega_1\tau})(e^{i\omega_2\tau})\nonumber \\
	=& \frac{(1-\alpha)\sigma}{\sqrt{32\pi^3}}\> e^{-2\sigma^2(\delta+\tau)^2}.
	\label{POVMs_eqn5}
\end{align}

Similarly, we evaluate
\begin{align}
	\bra{\psi}\hat{d}_1^\dagger(t_0)\hat{c}_2^\dagger(t_0+\tau)\ket{0} 
	&= \frac{1}{2}\bra{0}\int d\omega \> \phi(\omega) e^{-i\omega\delta} 
	(\sqrt{\alpha}[\hat{c}_2(\omega)]+\sqrt{1-\alpha}[-i\hat{d}_1(\omega)])[-i\hat{c}_2(\omegap-\omega)] 
	\hat{d}_1^\dagger(t_0)\hat{c}_2^\dagger(t_0+\tau)\ket{0}\nonumber \\
	&= -\frac{\sqrt{1-\alpha}}{4\pi}\int d\omega \> \phi(\omega) e^{-i\omega\delta} 
	e^{-i(\omegap-\omega)(t_0+\tau)}e^{-i\omega t_0}\nonumber \\
	&= -\frac{\sqrt{1-\alpha}}{4\pi}\int d\omega \> \phi(\omega) e^{-i\omega\delta} 
	e^{-i\omegap(t_0+\tau)}e^{i\omega\tau}.
	\label{POVMs_eqn6}
\end{align}
And therefore
\begin{align}
	\bra{\psi}\hat{d}_1^\dagger(t_0)\hat{c}_2^\dagger(t_0+\tau)\ket{0}\bra{0}\hat{c}_2(t_0+\tau)\hat{d}_1(t_0)\ket{\psi} 
	&= \frac{1-\alpha}{16\pi^2}\int d\omega_1 \> \int d\omega_2 \> \phi(\omega_1)\phi^*(\omega_2) e^{-i(\omega_1-\omega_2)\delta} 
	(e^{i\omega_1\tau})(e^{-i\omega_2\tau})\nonumber \\
	&= \frac{(1-\alpha)\sigma}{\sqrt{32\pi^3}}\> e^{-2\sigma^2(\delta-\tau)^2}.
	\label{POVMs_eqn7}
\end{align}

We can now construct Eq.~\eqref{PA_eqn24} with Eqs.~(\ref{PA_eqn21},~\ref{POVMs_eqn3},~\ref{POVMs_eqn5},~\ref{POVMs_eqn7}), noting that the remaining terms in \( \hat{\Pi}_{\mathrm{c}}(\tau) \) are Eqs.~(\ref{POVMs_eqn3},~\ref{POVMs_eqn5},~\ref{POVMs_eqn7}) with \( -\tau \).

\subsection{HOM bunching probability}
Deriving the fundamental bunching rate follows similarly.
We seek to evaluate ${P_\mathrm{b}(\tau)=\bra{\psi}\hat{\Pi}_\mathrm{b}(\tau)\ket{\psi}}$, with the bunching \acp{POVM} $\hat{\Pi}_\mathrm{b}(\tau)$ given in Eq.~\eqref{PA_eqn22}.
The key difference is we now look for cases where photons arrive at the same detector.

Again relabelling $t\to t_0 +\tau/2$. we start by evaluating
\begin{align}
	\bra{\psi}\hat{c}_1^\dagger(t_0)\hat{c}_1^\dagger(t_0+\tau)\ket{0} 
	=& \frac{1}{2}\bra{0}\int d\omega \> \phi(\omega) e^{-i\omega\delta} \sqrt{\alpha} 
	[-i\hat{c}_1(\omega)][\hat{c}_1(\omegap-\omega)] 
	\hat{c}_1^\dagger(t_0)\hat{c}_1^\dagger(t_0+\tau)\ket{0}\nonumber \\
	=& -i\frac{\sqrt{\alpha}}{4\pi}\int d\omega \> \phi(\omega) e^{-i\omega\delta} 
	(e^{-i\omega t_0}e^{-i(\omegap-\omega)(t_0+\tau)}+e^{-i(\omegap-\omega)t_0}e^{-i\omega(t_0+\tau)}) \nonumber \\
	=& -i\frac{\sqrt{\alpha}}{4\pi}e^{-i\omegap t_0}\int d\omega \> \phi(\omega) e^{-i\omega\delta} 
	(e^{-i\omega\tau}+e^{-i\omegap\tau}e^{i\omega \tau}).
	\label{POVMs_eqn8}
\end{align}
Multiplying by its conjugate, we obtain
\begin{align}
	\bra{\psi}\hat{c}_1^\dagger(t_0)\hat{c}_1^\dagger(t_0+\tau)\ket{0}\bra{0}\hat{c}_1(t_0+\tau)\hat{c}_1(t_0)\ket{\psi}
	=& \frac{\alpha}{16\pi^2}\int d\omega_1 \> \int d\omega_2 \> \phi(\omega_1)\phi^*(\omega_2) e^{-i(\omega_1-\omega_2)\delta} \nonumber\\
	&\mkern32mu\times (e^{-i\omega_1\tau}+e^{-i\omegap\tau}e^{i\omega_1 \tau})(e^{i\omega_2\tau}+e^{i\omegap\tau}e^{-i\omega_2 \tau})\nonumber \\
	=& \frac{\alpha\sigma}{\sqrt{32\pi^3}} \> e^{-2\sigma^2(\delta+\tau)^2}(e^{4\delta\sigma^2\tau}+1)^2
	\label{POVMs_eqn9}
\end{align}

For the next two terms, we note that that $\hat{c}^\dagger$ and $\hat{d}^\dagger$ modes are independent.
They do not interfere at the beamsplitter, therefore these terms depend only on arrival time and have the same probability as if they arrived at different detectors.
Therefore
\begin{align}
	\bra{\psi}\hat{c}_1^\dagger(t_0)\hat{d}_1^\dagger(t_0+\tau)\ket{0}\bra{0}\hat{d}_1(t_0+\tau)\hat{c}_1(t_0)\ket{\psi} 
	=&\bra{\psi}\hat{c}_1^\dagger(t_0)\hat{d}_2^\dagger(t_0+\tau)\ket{0}\bra{0}\hat{d}_2(t_0+\tau)\hat{c}_1(t_0)\ket{\psi} \nonumber \\*
	=& \frac{(1-\alpha)\sigma}{\sqrt{32\pi^3}}\> e^{-2\sigma^2(\delta+\tau)^2}
	\label{POVMs_eqn10}
\end{align}
and
\begin{align}
	\bra{\psi}\hat{d}_1^\dagger(t_0)\hat{c}_1^\dagger(t_0+\tau)\ket{0}\bra{0}\hat{c}_1(t_0+\tau)\hat{d}_1(t_0)\ket{\psi} 
	=&\bra{\psi}\hat{d}_1^\dagger(t_0)\hat{c}_2^\dagger(t_0+\tau)\ket{0}\bra{0}\hat{c}_2(t_0+\tau)\hat{d}_1(t_0)\ket{\psi} \nonumber \\*
	=& \frac{(1-\alpha)\sigma}{\sqrt{32\pi^3}}\> e^{-2\sigma^2(\delta-\tau)^2}.
	\label{POVMs_eqn11}
\end{align}

As before, we can now construct Eq.~\eqref{PA_eqn36} with Eqs.~(\ref{PA_eqn22}, \ref{POVMs_eqn9}, \ref{POVMs_eqn10}, \ref{POVMs_eqn11}), noting that the remaining terms in \( \hat{\Pi}_{\mathrm{b}}(\tau) \) are bunchings in the second detector which are identical due to the symmetry of the problem.

\subsection{No-HOM arrival probability}
For our no-HOM protocol, we are concerned only with arrival times.
For the average arrival time $t$, and difference in arrival times $\tau$, we have the \ac{POVM} element
\begin{equation}
\begin{aligned}
	\hat{\Pi}^{\mathrm{NH}}(\tau) = 2\pi\bigg[\hat{a}_i^\dagger\left(t-\frac{\tau}{2}\right)\hat{a}_s^\dagger\left(t+\frac{\tau}{2}\right)\ket{0}\bra{0}\hat{a}_s\left(t+\frac{\tau}{2}\right)\hat{a}_i\left(t-\frac{\tau}{2}\right)  + \hat{b}_i^\dagger\left(t-\frac{\tau}{2}\right)\hat{a}_s^\dagger\left(t+\frac{\tau}{2}\right)\ket{0}\bra{0}\hat{a}_s\left(t+\frac{\tau}{2}\right)\hat{b}_i\left(t-\frac{\tau}{2}\right)\bigg].
\end{aligned}
	\label{ID_eqn10}
\end{equation}
We recall that for this protocol, we allow for both positive and negative $\tau$. The factor of $2\pi$ again comes from our choice of a monovariate spectral distribution. We note that for strict completeness we also have the \ac{POVM} element \( \mathbb{1} - \int d\tau \> \hat{\Pi}^{\mathrm{NH}}(\tau) \), though we can again omit this from our calculation as this occurs with zero probability for states of form Eq.~\eqref{PA_eqn99}.

We want to fully derive $P^\mathrm{NH}(\tau)=\bra{\psi}\hat{\Pi}^{\mathrm{NH}}(\tau)\ket{\psi}$.
As we no longer have a beamsplitter present, the photons never interact after being generated and there is no dependence on their relative indistinguishability
as such we can leave the second mode as \( \hat{a}_{\mathrm{s}}^{\dagger} \) where--for the purposes of this detection scheme only---\( \hat{a}_{\mathrm{i}}^{\dagger} \) and \( \hat{a}_{\mathrm{s}}^{\dagger} \) need not be identical up to the spatial degree of freedom.
Thus we have the state
\begin{equation}
	\ket{\psi} = \int d\omega \> \phi(\omega) e^{i\omega\delta} \hat{a}_i^\dagger(\omega)\hat{a}_s^\dagger(\omegap-\omega)\ket{0}
	\label{ID_eqn9}
\end{equation}
and likewise use the simplified \ac{POVM} element
\begin{equation}
	\hat{\Pi}^{\mathrm{NH}}(\tau)= 2\pi \> \hat{a}_i^\dagger\left(t-\frac{\tau}{2}\right)\hat{a}_s^\dagger\left(t+\frac{\tau}{2}\right)\ket{0}\bra{0}\hat{a}_s\left(t+\frac{\tau}{2}\right)\hat{a}_i\left(t-\frac{\tau}{2}\right)
	\label{POVMs_eqn14}.
\end{equation}

Relabelling  $t\to t_0 +\tau/2$, we evaluate
\begin{align}
	\bra{\psi}\hat{a}_i^\dagger(t_0)\hat{a}_s^\dagger(t_0+\tau)\ket{0} 
	=&\bra{0}\int d\omega \> \phi(\omega) e^{-i\omega\delta} 
	\hat{a}_i^\dagger(\omega)\hat{a}_s^\dagger(\omegap-\omega)\hat{a}_i^\dagger(t_0)\hat{a}_s^\dagger(t_0+\tau)\ket{0} \nonumber \\*
	=& \frac{1}{2\pi} \int d\omega \> \phi(\omega) e^{-i\omega\delta} e^{-i(\omegap-\omega)(t_0+\tau)} e^{-i\omega t_0}\nonumber \\*
	=& \frac{1}{2\pi} \int d\omega \> \phi(\omega) e^{-i\omega\delta} e^{-i\omegap(t_0+\tau)} e^{i\omega \tau}.
	\label{POVMs_eqn12}
\end{align}
Multiplying by its conjugate we then obtain
\begin{align}
	\bra{\psi}\hat{a}_i^\dagger(t_0)\hat{a}_s^\dagger(t_0+\tau)\ket{0}\bra{0}\hat{a}_s(t_0+\tau)\hat{a}_i(t_0)\ket{\psi} 
	&=\int d\omega_1 \> \int d\omega_2 \> \phi(\omega_1)\phi^*(\omega_2) e^{-i(\omega_1-\omega_2)\delta} 
	(e^{i\omega_1\tau})(e^{-i\omega_2\tau})\nonumber \\
	&=\frac{1}{\sqrt{2\pi^3}} \sigma e^{-2\sigma^2(\delta-\tau)^2},
	\label{POVMs_eqn13}
\end{align}
which, combined with Eq.~\eqref{POVMs_eqn14}, gives Eq.~\eqref{ID_eqn7}.

\section{Fundamental limits}\label{qfiCalcs}

For the \ac{QFI} calculation we must evaluate the quantities $\braket{\partial_{\delta} \psi | \psi }$ and $\braket{\partial_{\delta} \psi | \partial_{\delta} \psi}$.
We first evaluate the overlap $\braket{\partial_\delta\psi|\psi}$:
\begin{align}
	\braket{\partial_\delta\psi|\psi} &=\frac{1}{V} \int d\omega_1 \int d\omega_2\> \phi(\omega_1)\phi(\omega_2)(-i\omega_1)) 
	e^{-i\omega_1\delta}e^{i\omega_2\delta} \nonumber \\*
	& \qquad \qquad \qquad \times (\alpha\bra{0}\hat{a}_s(\omegap-\omega_1)\hat{a}_i(\omega_1) 
	\hat{a}_i^\dagger(\omega_2)\hat{a}_s^\dagger(\omegap-\omega_2)\ket{0} \nonumber \\*
	& \qquad \qquad \qquad \qquad + (1-\alpha)\bra{0}\hat{a}_s(\omegap-\omega_1)\hat{b}_i(\omega_1) 
	\hat{b}_i^\dagger(\omega_2)\hat{a}_s^\dagger(\omegap-\omega_2)\ket{0}) \nonumber \\
	&= -\frac{i}{V} \int d\omega_1 \int d\omega_2\> \phi(\omega_1)\phi(\omega_2)\omega_1 
	e^{-i\omega_1\delta}e^{i\omega_2\delta} 
	\delta_\mathrm{D}((\omegap-\omega_2)-(\omegap-\omega_1))\delta_\mathrm{D}(\omega_2-\omega_1) \nonumber \\
	&= -\frac{i}{V} \int d\omega \> \phi(\omega)^2 (\omegap-\omega) \delta_\mathrm{D}(0) \nonumber \\*
	&= -\frac{i\omegap}{2}.
\end{align}

We can similarly evaluate the overlap $\braket{\partial_\delta\psi|\partial_\delta\psi}$ to find
\begin{equation}
	\braket{\partial_\delta\psi|\partial_\delta\psi} = \frac{1}{V} \int d\omega \> \phi(\omega)^2 \omega^2 \delta_\mathrm{D}(0) 
	= \sigma^2 + \frac{\omegap^2}{4}.
\end{equation}

We then combine these with Eq.~\eqref{ID_eqn2} to obtain the \ac{QFI} as given in Eq.~\eqref{ID_eqn5}.

\vspace{0.25cm}
\pagebreak

\twocolumngrid

\section{Fisher information matrices without time-resolution}\label{no_TR_FIMs}

Without time-resolution we have a finite number of measurement outcomes.
It is therefore simple to show the full form of the \ac{FIM}, for both detector types.

\subsection{Bucket detectors}
Let $\vec{\theta}=(\delta,\alpha,\sigma,\gamma)$ be the vector of potentially unknown parameters.
Then for bucket detectors without time-resolution, with elements as defined in Eq.~\eqref{FI_eqn1}, our \ac{FIM} takes the form
\begin{equation}
	\vec{F}^\mathrm{B}(\vec{\theta}) =
	\begin{pmatrix}
		 16 \alpha^2 \delta^2 \kappa \sigma^4 & -4 \alpha \delta \kappa \sigma^2 & 16 \alpha^2 \delta^3 \kappa \sigma^3 & 8 \alpha \delta \sigma^2 \chi \\
		 -4 \alpha \delta \kappa \sigma^2 & \kappa & -4 \alpha \delta^2 \kappa \sigma & -2 \chi \\
		 16 \alpha^2 \delta^3 \kappa \sigma^3 & -4 \alpha \delta^2 \kappa \sigma & 16 \alpha^2 \delta^4 \kappa \sigma^2 & 8 \alpha \delta^2 \sigma \chi \\
		 8 \alpha \delta \sigma^2 \chi & -2 \chi & 8 \alpha \delta^2 \sigma \chi & -\frac{8 \chi e^{2 \delta^2 \sigma^2} }{(1- \gamma)^2}
	\end{pmatrix}
	,
\label{no_TR_FIMs_eqn1}
\end{equation}
with
\begin{align}
	\kappa &= \frac{(1-\gamma )^2 (1+\gamma)}{\alpha^2 (\gamma -1)-4 \alpha \gamma e^{2 \delta^2 \sigma^2}+(3 \gamma +1) e^{4 \delta^2 \sigma^2}}, \nonumber \\*
	\chi &= \frac{1-\gamma }{\alpha (\gamma -1)-(3 \gamma +1) e^{2 \delta^2 \sigma^2}}.
	\label{no_TR_FIMs_eqn2}
\end{align}
This matrix is rank 2, therefore it is singular and a multiparameter estimation of all four parameters is impossible.
Only submatrices covering exactly one of $\{\delta,\alpha,\sigma\}$ along with $\gamma$ are non-singular.
It is therefore possible to estimate the photon loss at the same time as estimating any one of the other parameters.
The singularity of this matrix is removed through the introduction of time-resolution.

\subsection{Number-resolving detectors}
For number-resolving detectors without time-resolution, our FIM now takes the form
\begin{equation}
	\vec{F}^\mathrm{NR}(\vec{\theta}) =
\begin{pmatrix}
 16 \alpha^2 \delta^2 \xi  \sigma^4 & -4 \alpha  \delta  \xi  \sigma^2 & 16 \alpha^2 \delta^3 \xi  \sigma^3 & 0 \\
 -4 \alpha  \delta  \xi  \sigma^2 & \xi  & -4 \alpha  \delta^2 \xi  \sigma  & 0 \\
 16 \alpha^2 \delta^3 \xi  \sigma^3 & -4 \alpha  \delta^2 \xi  \sigma  & 16 \alpha^2 \delta^4 \xi  \sigma^2 & 0 \\
 0 & 0 & 0 & \frac{2}{\gamma -\gamma^2}
\end{pmatrix} ,
\label{no_TR_FIMs_eqn3}
\end{equation}
with
\begin{equation}
	\xi = \frac{(1-\gamma )^2}{e^{4 \delta^2 \sigma^2}-\alpha^2}.
	\label{no_TR_FIMs_eqn4}
\end{equation}

Once again, this matrix is rank 2.
The key difference here is that by eliminating the bunching/loss ambiguity $\gamma$ is now wholly independent, this manifests in the FIM by setting all off-diagonal terms involving $\gamma$ to zero.
This means that $\gamma$ can be estimated even if none of the other parameters are known, the estimation will be just as efficient regardless of what the other parameters are set to, and estimating $\gamma$ simultaneously with any other parameters will not harm the efficiency of the other estimations.
As before the singularity vanishes once time-resolution is introduced, though there is still some correlation between $\delta$, $\alpha$, and $\sigma$.

\subsection{Perfect visibility}
\begin{figure}[hbt]
\centering
\includegraphics[width=\linewidth]{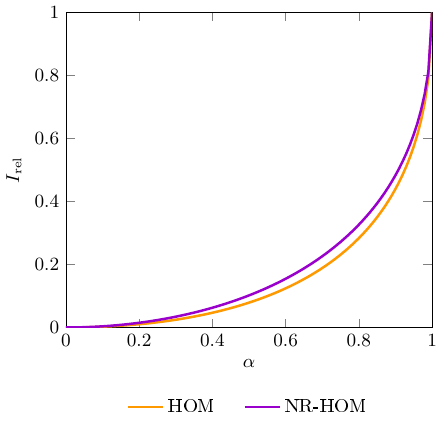}
\caption{The relative information $\Irel$ for our non-time-resolving protocols.
We set $\gamma = 0.4$, choose $\delta$ to maximise information, and vary $\alpha$.
As we approach $\alpha=1$, the two information peaks converge, leading to a single peak at $\delta=0$ where $F_\delta=\Htwoph$ and we have reached the QFI limit.}
\label{appB_fig}
\end{figure}

From Eqs.~\eqref{no_TR_FIMs_eqn1} and~\eqref{no_TR_FIMs_eqn2}, we see for bucket HOM the single-parameter \ac{CFI} $F^\mathrm{B}_\delta=[\vec{F}^\mathrm{B}(\vec{\theta})]_{1,1}$ takes the form
\begin{equation}
	F^\mathrm{B}_\delta = \frac{16\alpha^2(1-\gamma )^2 (\gamma +1)\delta^2\sigma^4}{\alpha^2 (\gamma -1)-4 \alpha  \gamma  e^{2 \delta^2 \sigma^2}+(3 \gamma +1) e^{4 \delta^2 \sigma^2}}.
	\label{no_TR_FIMs_eqn5}
\end{equation}
As we approach perfect visibility, $\alpha\to1$, and in the limit $\delta \to 0$, we obtain $F^\mathrm{B}_\delta\to4\sigma^2(1-\gamma)^2=\Htwoph(\gamma)$, the two-photon conditioned \ac{QFI}.
Similarly from Eqs.~\eqref{no_TR_FIMs_eqn3} and \eqref{no_TR_FIMs_eqn4} the NR HOM single-parameter \ac{CFI} $F^\mathrm{NR}_\delta=[\vec{F}^\mathrm{NR}(\vec{\theta})]_{1,1}$ takes the form
\begin{equation}
	F^\mathrm{NR}_\delta = \frac{16\alpha^2(1-\gamma )^2\delta^2\sigma^4}{e^{4 \delta^2 \sigma^2}-\alpha^2}.
	\label{no_TR_FIMs_eqn6}
\end{equation}
Like with bucket detectors, by letting $\alpha\to1$ and $\delta\to0$, we find $F^\mathrm{NR}_\delta\to4\sigma^2(1-\gamma)^2=\Htwoph(\gamma)$, and have again recovered the two-photon conditioned \ac{QFI}.
The maximal information as a function of $\alpha$ is plotted in Fig.~\ref{appB_fig}.

\section{Estimating other parameters}\label{other_paras}

We have throughout this paper focused on estimation of the delay $\delta$.
A key merit of our time-resolving HOM protocols is that they allow $\delta$ to be estimated even when other parameters are unknown, streamlining the calibration process compared to previous protocols.
We now briefly examine the \ac{CFI} for these other parameters and discuss how they might best be estimated.
\begin{figure}[htbp]
\centering
\includegraphics[width=\linewidth]{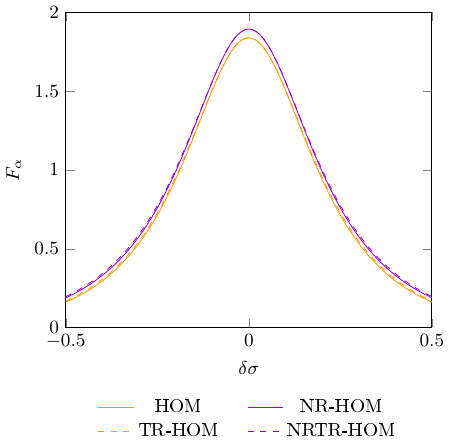}
\caption{Comparison of the \ac{CFI} $F_\alpha$ for our various protocol configurations, as we vary the delay $\delta$. We set $\alpha = 0.9, \gamma = 0.4, T = 5/\sigma$. The information peaks at $\delta = 0$, where coincidence and bunching rates vary most rapidly with $\alpha$. The benefits of time-resolution are small, both bucket curves closely overlap, as do both number-resolving curves, and at $\delta=0$ the difference in information between time-resolving and non-time-resolving protocols vanishes.}
\label{appC_fig3}
\end{figure}
We discussed in Sec.~\ref{sec_results} how estimating both $\delta$ and $\alpha$ simultaneously shifted the optimal estimation points closer to the origin compared to those when just estimating $\delta$.
We can here see, in Fig.~\ref{appC_fig3}, that the \ac{CFI} $F_\alpha$ indeed peaks exactly at the origin, this being the optimal place to estimate $\alpha$ alone as here the coincidence and bunching rates are most sensitive to changes in the visibility.

\begin{figure}[htbp]
\centering
\includegraphics[width=\linewidth]{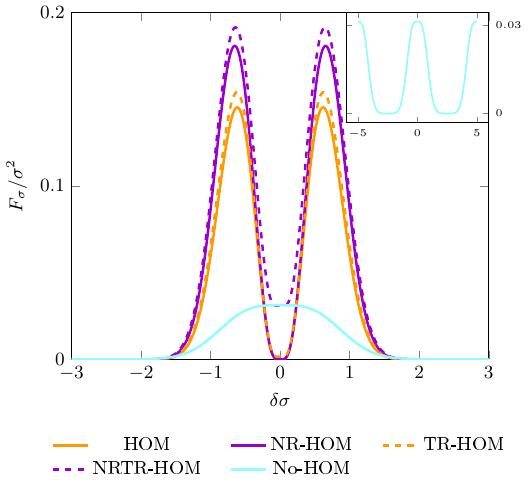}
\caption{Comparison of the \ac{CFI} $F_\sigma$ for our various protocol configurations, as we vary the delay $\delta$.
We set $\alpha = 0.9$, $\gamma = 0.4$, $T = 5/\sigma$.
These follow a similar shape to the plots of $F_\delta$ in Fig.~\ref{results_fig1}; the main differences being that the peaks are narrower, more spread out, and for TR-HOM and NRTR-HOM $F_\sigma$ does not dip all the way to zero at the origin.
With the inset we can see that the no-HOM information follows a similar oscillatory pattern, with maxima and minima in the same positions but an overall more squared curve.}
\label{appC_fig1}
\end{figure}

We see in Fig.~\ref{appC_fig1} how the \ac{CFI} $F_\sigma$ varies with $\delta$.
When all other parameters are known, the optimum position to estimate $\sigma$ is with $\delta$ at either of the peaks.
Suppose that we want to estimate $\sigma$ simultaneously with both $\delta$ and $\alpha$. The peaks in $F_\sigma$ lie further from the origin compared to the peaks for $F_\delta$.
While estimating $\alpha$ with $\delta$ pulls the optimum estimation points inward, the additional requirement of estimating $\sigma$ will again pull them back outwards.

For $\gamma$, we previously noted in Appendix~\ref{no_TR_FIMs} that when number-resolving detectors are used this becomes an independent parameter, and can be estimated trivially without impacting the estimation of any other parameters.
When using bucket detectors, we treated $\gamma$ as a calibration parameter, obtained very simply by tuning ourselves far outside the dip.
Even though $\gamma$ is not independent in this case, it can still be estimated simultaneously with other parameters.
$F_\gamma$ is also independent of any time-resolution for our detectors.

\begin{figure}[htbp]
\centering
\includegraphics[width=\linewidth]{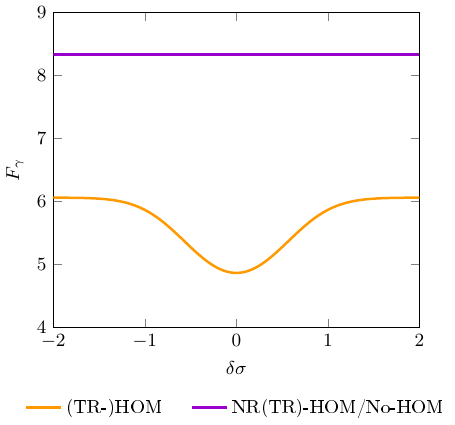}
\caption{Comparison of the \ac{CFI} $F_\gamma$ for our various protocol configurations, as we vary the delay $\delta$.
We set $\alpha = 0.9$ and $\gamma = 0.4$.
No-HOM is functionally identical to NR(TR)-HOM in regards to $F_\gamma$, as the photons are always directed to separate detectors. In these cases the information is constant.
For (TR-)HOM, the information is reduced somewhat due to the bunching/loss ambiguity. It tends to a constant for large $\delta$, but further dips slightly near the origin where bunching is more likely.
}
\label{appC_fig2}
\end{figure}

In Fig.~\ref{appC_fig2} we see a constant $F_\gamma$ when number-resolving detectors are used.
This is also the case for our no-HOM protocol, as each photon always arrives at a different detector.
Bucket detectors perform slightly worse, and we see an additional small dip near the origin.
There is no dependence on time-resolution.
Therefore, when using bucket detectors for our HOM protocol, a requirement to estimate $\gamma$ with the other parameters would further push the optimal estimation point outwards once again.

\end{document}